\documentclass[dvipsnames]{article}
\usepackage[UKenglish]{babel}
\usepackage{array}
\usepackage{url}
\usepackage{multirow}
\PassOptionsToPackage{printonlyused,smaller}{acronym}
  \usepackage{acronym}
\pdfpageheight\paperheight
\pdfpagewidth\paperwidth

\providecommand{\tabularnewline}{\\}

\usepackage[section]{placeins} %

\usepackage{colortbl}
\usepackage{pifont} %
\usepackage{tikz}
\newcommand\crossout[2]{%
  \multicolumn{1}{p{#2}|}{
  \hskip-\tabcolsep
  $\vcenter{\begin{tikzpicture}[baseline=0,anchor=south west,inner sep=#1]
  \path[use as bounding box] (0,0) rectangle (#2+2\tabcolsep,\baselineskip);
  \node[minimum width={#2+2\tabcolsep},minimum height=\baselineskip+\extrarowheight] (box) {};
  \draw (box.north west) -- (box.south east) (box.north east) -- (box.south west);
 \end{tikzpicture}}$\hskip-\tabcolsep
}}
\newcommand\xoutnovcm[2]{%
  \multicolumn{1}{p{#2}}{
  \hskip-\tabcolsep
  $\vcenter{\begin{tikzpicture}[baseline=0,anchor=south west,inner sep=#1]
  \path[use as bounding box] (0,0) rectangle (#2+2\tabcolsep,\baselineskip);
  \node[minimum width={#2+2\tabcolsep},minimum height=\baselineskip+\extrarowheight] (box) {};
  \draw (box.north west) -- (box.south east) (box.north east) -- (box.south west);
 \end{tikzpicture}}$\hskip-\tabcolsep
}}

\usepackage{caption} %
\usepackage{changepage}
\usepackage{ragged2e} %
\usepackage{multicol}

\PassOptionsToPackage{%
  backend=bibtex8,bibencoding=ascii,%
  language=auto,%
  style=authoryear-comp, %
  maxcitenames=2, %
  mincitenames=1,  
  maxbibnames=8, %
  minbibnames=1, %
  sorting=nyt, %
  sortcites=true, %
  abbreviate=true, %
  natbib=true %
}{biblatex}
    \usepackage{biblatex}

\makeatother

\begin{document}
\date{April 2018}

\title{Review of Human Decision-making during Incident Analysis}

\author{Jonathan M. Spring, Phyllis Illari}

\maketitle

\begin{abstract}
We review practical advice on decision-making during computer security incident response. 
Scope includes standards from the IETF, ISO, FIRST, and the US intelligence community.
To focus on human decision-making, the scope is the evidence collection, analysis, and reporting phases of response.
The results indicate both strengths and gaps. 
A strength is available advice on how to accomplish many specific tasks.
However, there is little guidance on how to prioritize tasks in limited time or how to interpret, generalize, and convincingly report results.
Future work should focus on these gaps in explication and specification of decision-making during incident analysis. 
\end{abstract}

\section{Introduction}
\label{sec:IR-intro}

The purpose of this literature review is to identify the structure of human decision-making during computer-security incident response (shortened as incident response if the usage is unambiguous) and identify the existing research programs that underpin the study of each element of the decision-making structure. The utility of such a literature review is to identify opportunities to expand research programs, as well as to identify loci of research programs best strengthened by interdisciplinary work or requiring interfield theories. 

Incident response is an attractive topic because it anchors the whole field of information security. When information security is not responding to an incident, it is either preparing for one or learning from past incident responses.\footnote{Prevention of future incidents may fruitfully be discussed as independent from incident response, but in practice prevention techniques are almost all adapted from the lessons learned after responding to an incident.} Preparation and learning are each complex and independent. However, their crux is incident response. We restrict our focus to human-centric decision making as opposed to automated or machine-learning techniques.  Such automation is clearly crucial to incident response. However, deciding what automation to build and use remains a human decision, as does how to interpret and act on results. 

Information security is important, yet breaches are usually detected months after they occur~\citep{verizon2015dbir}. Professionals suggest judging whether an organization has a good security posture by their incident response capabilities \citep[ch. 15]{shimeall2013introduction}.\footnote{See also: ``The question you must accept is not whether security incidents will occur, but rather how quickly they can be identified and resolved.'' \url{https://www.gartner.com/smarterwithgartner/prepare-for-the-inevitable-security-incident/}} The question we address in this review is: \emph{how do incident responders make decisions during an investigation?} 

Successful guidance on this question improves incident response, and thereby all of information security. However, the question is deceptively simple, and the use of `how' is ambiguous. More specifically, we want a satisfactory process by which incident responders generate and prioritize questions to ask, a satisfactory process of answering these questions under time constraints, and a satisfactory process for how these answers lead to decisions. In summary, there is ample advice on what questions to ask and how to answer a given specific question, but little on how to prioritize questions or interpret results. Therefore, we identify a significant lacuna in the existing literature on decision making and evidence evaluation during incident response. 
We suggest some directions for filling this gap use mathematical modeling and philosophy of science. 

The immediate challenge of this literature review is an explosion of scope. The scope is too broad for at least two reasons: the topic is of broad application with many sub-parts, and the academic literature is not the only relevant source. Practitioners are a necessary source if the review is to adequately capture the state of the art. Besides simply increasing the volume of publications to review, information security practitioners do not publish with the same reliability nor norms as customary for input to an academic literature review. This variability makes it difficult to evaluate contributions on a single appraisal strategy. 

We approach this problem in the two broad parts. The first part covers scope and problem definition. Section~\ref{sec:Scope-Topic} defines the term computer-security incident response, and which aspects are in scope for our discussion. Section~\ref{sec:Scope-Publication-Venues} defines the scope in terms of publication venues included in the review.  Section~\ref{sec:Novelty} provides evidence that no literature review with similar scope and goals has been conducted. Section~\ref{sec:Methods} defines our method for reviewing the state of decision-making recommendations during computer-security incident response. 

The second part covers results and proposed research. Section~\ref{sec:Results} provides the results of the incident response decision-making literature review. The first four subparts cover each of the four venues in scope, and Section~\ref{sub:results-Referenced-Documents} reports the evaluation of documents cited by those documents from the search results that are evaluated to be relevant. 
Section~\ref{sec:Discussion} critically analyzes the results of the reviews. 
Section~\ref{sec:IR-conclude} suggests gaps and some broad directions that might be followed to begin to fill them.

\section{Scope\label{sec:Scope}}

This section limits the scope of the literature review in two distinct aspects: the definition of the topic and the publication venues. 
We will restrict the definition of \ac{CSIR} to three subtasks during investigation: evidence collection, analysis, and reporting. 
We will restrict publication venues to rele\-vant international standards and academic literature that is referenced therein. Specifically, the search venues will be the \acf{IETF}, the \acf{ISO},\footnote{Although \ac{ISO} standards are only available for a fee, the terms and definitions as used in the rele\-vant standards (the 27000 series) are freely available.}  \acf{FIRST}, and documents understood to represent the \ac{US} \acf{IC}. 

As far as possible, we will use standard definitions for terms, preferring global, consensus, freely-available definitions. Explicitly, in order of preference, the \ac{IETF}, \ac{ISO}, and \ac{FIRST}. This ordering is based on the extent of consensus (the \ac{IETF} membership is broader than \ac{FIRST}) and the openness of the definitions. Otherwise, the choice of established definitions for jargon is primarily for clarity, and to compress the discussion; we assume familiarity with the terms in the \ac{IETF} Internet Security Glossary \citep{rfc4949}. 

The scope is not the traditional academic venues. 
The operational reason is to focus on what incident responders actually do. 
Ideally this would take the form of first-hand accounts; however, cybersecurity is a sensitive topic. Chatham House Rules\footnote{Chatham House Rules indicates a situation in which the information or content of a meeting, discussion, or presentation may be disclosed but the source of the information may not be identified, implicitly or explicitly. This request is made by the speaker prior to disclosing the information. } and \acp{NDA} abound in the discipline; these norms within the security community further frustrate any usual academic publication expectations. 
It would be impossible to evaluate the selection bias or outright deception within studies. 
The incident response standards at least form an honest baseline of what is expected of a competent practitioner. 
The assumption is that this review applies to competent practitioners, where competent is defined by consensus among practitioners and codified in the standards. 
We do not empirically address the extent to which competence is common.  
Due to the community norms of secrecy, norms documented by \citet{sundaramurthy2014anthropological}, a comprehensive evaluation seems impractical. 

Section~\ref{sec:Novelty} provides evidence that the academic literature does not systematically cover our scope. 
The method to justify this is a search through two common sources of literature reviews, \ac{ACM} Computing Surveys and the \ac{IEEE} Security and Privacy ``systematization of knowledge'' papers.
In summary, no relevant surveys have been published on human decision-making during \ac{CSIR} analysis.  
While the reason why is unclear, the task is not made easier by the natural secrecy of practitioners. 

A further consideration in focusing on standards as the starting point is simply to prevent an explosion of documents to review. A cursory Google Scholar search for ``computer security incident response'' and ``digital forensic investigation'' each return about 2,000 results.  Alternatively, searches in the \ac{ACM} Guide to Computing Literature and \ac{IEEE} Xplore databases for ``computer security incident response'' each return 20-25 results (both on Sep 1, 2017). 
Many of these results are obscure, and for those that are not it is challenging to evaluate their operational impact. 
Going straight to the standards bypasses this question of impact -- the remit and authority of standards is explicit. 

Finally, one author draws on six years of personal experience working at a research center interacting with a variety of practicing incident responders. %

\subsection{Scope---Topic\label{sec:Scope-Topic}}

The first task is to situate computer-security incident response within a context of what it excludes and what falls under it. Incident response is a subspecies of business continuity planning or continuity of operations.  Continuity planning may be a response to man-made or natural events, and either physical or digital events. A military invasion is a man-made, physical event; a hurricane is a natural, physical event. Computer-security incident response is only in response to security incidents that are primarily digital, where a security incident is something ``contrary to system policy'' \citep{rfc4949}. Thus, accidents of all kinds are out of scope; though distinguishing apparent accidents from malicious acts is included. Intentional physical destruction of computing resources is also out of scope of computer-security incident response \citep{rfc2350}. 

Narrowing the focus further, incident response is a task within incident management.\footnote{The term ``incident management'' does not appear in \ac{IETF} documents consistently. \citet{rfc6684} describes  \ac{IODEF} \citep{rfc5070} as a protocol for ``exchange of incident management data,'' but the term ``incident management'' does not appear again in \citet{rfc6684}, and not even once in \citet{rfc5070}.  \citet{iso27035-1-2016} defines ``information security incident management'' as ``exercise of a consistent and effective approach to the handling of information security incidents.'' \ac{FIRST} does not provide a definition itself, but \citet{FIRST2017index} recommends the \ac{CERT/CC} documentation on incident management.
Trammell and Danyliw both worked at \ac{CERT/CC}, so this is probably the source of the informal reference in the \ac{IETF} documents.  The \ac{CERT/CC} phases are consistent with the \citet{iso27035-1-2016} phases of plan and prepare; detection and reporting; assessment and decision; responses; and lessons learnt. We prefer the \ac{CERT/CC} definitions as they are public (vice the \ac{ISO} standards), and recommended by \ac{FIRST} (thus in scope of using global, consensus-driven definitions).} \ac{CERT/CC} definitions of incident management \citep{alberts2004defining,mundie2014incident} locate incident response as independent from activities such as preparation, improving defences, training, financing, and lessons learnt. \citet{mundie2014incident} surveys practices including those by \ac{CERT/CC} and \ac{ISO}; six tasks are included as part of incident response: monitoring, detection, evidence collection, analysis, reporting, and recovery. These six tasks form the core topic of this survey of incident response. 

The human-centric decisions that are elements of these six incident response tasks vary in importance. Analysis, reporting, and recovery are almost wholly human-driven. Monitoring is almost wholly automated, while detection and evidence collection are a mixture. Where detection is automated, say in an \ac{IDS},\footnote{\citet{rfc4949} refers to \citet{bace2001ids} for the definition and guidelines for \ac{IDS}, which has been superseded by \citet{nist800-94}.} it is out of scope. Decisions about what detection rules to implement in an \ac{IDS} are part of the preparation or improving defences phases of incident management, as a result of lessons learnt, and thus are also out of scope. Actual human intrusion detection is rare, and when it occurs is usually the result of analysis during incident response to some other, automatically-detected incident. Therefore, the focus on human-driven incident response investigation excludes monitoring and detection. 

The \ac{IETF} \citep{rfc4949,rfc2350} and \ac{CERT/CC} define neither ``investigation'' nor ``forensics'' in relation to the incident management process.  \citet{iso27043-2015} places investigation as the centrepiece of incident management, where the principles of incident management are to ``give guidance on the investigation of, and preparation to investigate, information security incidents'' \citet[\S 0]{iso27035-1-2016}.  In this way, \ac{ISO} uses ``investigation'' as a near-synonym to ``response'' in the \ac{IETF} and \ac{FIRST} literature. 

The use of `incident' emphasizes that the investigation is oriented towards the violation of some policy, possibly but not necessarily a law. Thus, modelling or analysing online crime is an investigation, and so is IT staff looking into a usage policy violation. Incident response or investigation is entwined with cybersecurity because one essential aspect of a defence strategy is feedback from investigation to `preparation' and `protection'~\citep{alberts2004defining}. 
However, detailed discussion of preparation and protection is placed out of scope.

Incident response, per \ac{IETF} and \ac{FIRST}, explicitly includes remediation.  
\ac{ISO} \citep{iso27043-2015} treats remediation and response as separate from investigation. 
In determining scope, we follow \ac{ISO} and exclude recovery. Note that both sets of standards agree that clear reporting is the proper output of incident analysis, and any recovery follows reporting. 
However, it does seem clear that recovery follows a different decision process than analysis, and the two should be treated separately.
Within the six tasks identified within incident response, three are left in scope:
\begin{itemize}
\item evidence collection
\item analysis
\item reporting
\end{itemize}
These three seem too tightly coupled to separate, and are described consistently across the international standards organizations.

For each of these three topics, the concern is primarily with how an individual analyst makes decisions during these three phases.  
What tool or what language the analyst or investigator uses to make these choices is not germane and is out of scope. 
This is not a review of available security tools, platforms, or data exchange formats. 
The goal is to survey how analysts enumerate options, evaluate choices, generalize results, and justify these steps.

\subsection{Scope---Publication Venues\label{sec:Scope-Publication-Venues}}

Incident response and investigation includes professional and business aspects; therefore the scope of viable sources incident response practices cannot justifiably be limited to academic sources. 
As \citet{spring2017why} documents, the science of security is an unsettled area of research rather than an area with anything like standards.
In fact, traditional academic publication venues contain little if anything about day-to-day incident response practices; academics do not do incident response themselves.
\citet{sundaramurthy2014anthropological} seems to mark the first anthropological study of a \ac{CSIRT}\footnote{\ac{CSIRT} is the general term, and will be used unless referring to a specific organization.} members and their attitudes, but this literature is not about the actual process of incident response; that is covered in the professional literature.

Therefore, to understand current incident response practices the scope of the review is internationally-relevant standards and whatever literature is referenced therein. The history of standards as its own industry is complex in its own right \citep{spring2011learned}. 
The Internet and IT standards are formed by heterogeneous processes by a wide variety of actors \citep{oksala1996structure}. 
Security-relevant standards are beginning to be seen as having their own unique requirements, distinct from IT standards generally \citep{kuhlman2016trust}. 
However, it is a separate project to analyse how \ac{CSIR} standards have come to be. 
The standards in this review are taken as-is, with the understanding that any interpretations should be made cautiously because the standards may not cleanly fit in to existing studies of how and why other IT standards are created. 
More than other IT standards, \ac{CSIR} standards are likely a codification of tacit practitioner knowledge \citep{nightingale2009tacit}.

The scope of publication venues is limited to \ac{ISO}, \ac{IETF}, \ac{FIRST}, and the \ac{US} \acf{IC}. This choice is based on what organizations are relevant in influencing or describing international incident response practices, which is in turn based in the history of the governance of the Internet. 
We mitigate potential over-restriction of focus by including any documents cited by standards publications. 
The reasoning for selecting these organizations specifically is as follows.

\ac{ISO} and the \ac{ITU} are the authoritative technology standards makers \citep[p. 11]{oksala1996structure}. 
The US federal government plays a dominant role in Internet development and standards, through the original \ac{ARPA} development under the \ac{DoD} and subsequent stewardship under the Department of Commerce.\footnote{Two important sub-parts of Commerce are Internet governance by the National Telecommunications and Information Administration and standards by \acf{NIST}.}  

\ac{ISO} is \emph{de dicto} where one looks for international standards.
Each nation-state is allowed to have one member is \ac{ISO}, namely the official national standards body representing all the industry-based bodies in each country. It is a federation of federations, representing a multitude of industries. \ac{ISO} standardizes things like the two-letter country codes (which have been adopted as \ac{DNS} top-level domains), paper sizes, and credit cards. The \ac{ITU} and their CIRT program\footnote{\url{http://www.itu.int/en/ITU-D/Cybersecurity/Pages/Organizational-Structures.aspx}} seems promising in name; however, their website publishes little besides an events list. It appears that content is provided by \ac{FIRST} members, companies, or other consultancies; the \ac{ITU} does not produce its own incident response materials or standards. This leaves only \ac{ISO} in scope of the potential authoritative international standards bodies. 

On the other hand, the \ac{IETF} is the \emph{de facto} place to go for international Internet standards because, for all intents and purposes, its standards are the Internet. The \ac{IETF} ``doesn't recognize kings---only running code'' and creates more pragmatic, open (freely-available) standards \citep[p. 12]{oksala1996structure}. Open standards happen to have won out on the Internet; \ac{IETF} standards like \ac{TCP/IP}, \ac{DNS}, and \ac{BGP} underpin every Internet connection. For a background history of how the precursor to the \ac{IETF} came to this dominant role, see \citet{hafner1998wizards}.
The other main open-standards body is the \ac{W3C}, which standardizes \ac{HTTP} and \ac{XML}, for example. 
\ac{W3C} stays narrowly focused on web standards, and although this includes important web security considerations, \ac{W3C} does not work on incident management, so we mark the group as out of scope. 

\ac{FIRST} is not part of this longer \ac{ICT} standards history. It was formed in 1990 specifically to coordinate among and represent the interests of \acp{CSIRT} globally. \ac{FIRST}'s mission includes developing and sharing best practices, as well as creating and expanding incident response teams \citep{firstmission}. \ac{FIRST} is the one and only global organization representing those who do human-centric incident response tasks. \ac{FIRST}'s work with \ac{UN} agencies like the \ac{ITU} also testifies to its global influence.
It is naturally included as in-scope. 

There are three organizations one might consider naturally in scope that are excluded. These are \ac{ENISA}, \ac{NIST}, and the \ac{US} \ac{DoD}. However, within the gray area between \ac{NIST} and the \ac{US} \acl{IC}, we identify a fourth set of \emph{de facto} standards. 

\ac{ENISA} is specifically focused on \acp{CSIRT} and information security.
The \ac{EU} makes available an independent evaluation of \ac{ENISA}'s limited activities.\footnote{``Annual ex-post evaluation of \ac{ENISA} activities'' \url{https://www.enisa.europa.eu/about-enisa/annual-ex-post-evaluation-of-enisa-activities}} Its function is coordination and capacity building among \acs{EU}-member \acp{CSIRT} and to provide advice on some \ac{EU} policies. While \ac{EU} directive 2016/1148 will increase \ac{ENISA}'s statutory powers when it comes into effect in November 2018, at this point \ac{ENISA} has little authority to force member states to take its advice. In the scheme of incident response practices, \ac{ENISA} -- founded in 2004 -- is quite young. \ac{ENISA} documents are informational, the one document interfacing with \ac{EU} standards is an extended definition of the term ``cybersecurity'' and what \ac{EU} work is done related to it \citep{enisa2015definition}. Oddly, the document does not even mention incident response as an area within cybersecurity,\footnote{Despite the fact that \ac{ENISA} sponsored an informational document on evidence gathering aimed at increasing understanding between \acp{CSIRT} and law enforcement \citep{enisa2015evidence}. } so it seems safe to leave \ac{ENISA} out of scope. 

\ac{NIST} is a difficult organization to place in or out of scope. It is part of the Department of Commerce, and so has loose ties to the remaining Internet stewardship invested in the National Telecommunications and Information Administration. Strictly, \ac{NIST} merely sets policies for how the US federal civilian government secures its IT infrastructure and responds to incidents \citep{cichonski2012_sp800-61}. This \ac{FISMA} policy responsibility is a part of \ac{NIST}'s larger role of ``advancing measurement science, standards, and technology in ways that enhance economic security and improve our quality of life'' \citep{nistmission}. Through this role, \ac{NIST} standardized \acsu{AES}, which is the \emph{de facto} global standard encryption algorithm. \ac{NIST} documents and standards are also cited by the \ac{IETF}, \ac{ISO}, and \ac{FIRST}, elevating certain \ac{NIST} work from a national to international status.  
We shall consider \ac{NIST} generally out of scope; however, many \ac{NIST} publications will be considered as works cited by the international standards organizations. 

There are two US federal government units that do not fall under \ac{NIST}'s authority -- the \ac{DoD} and the \ac{CIA}. These two organizations have not published incident response standards as openly as \ac{NIST} or these other standards organizations. The \ac{DoD} does have other cybersecurity documents that are tangentially relevant. The Orange Book \citep{brand1985dod}, which evaluates trusted computing platforms, is relevant background material for incident responders. 

The questions the \ac{DoD} and its sub-agency the \ac{NSA} have raised around whether cybersecurity is, broadly, a science (see, e.g., \citet{mitre2010cyberscience,galison2012cybersecurity,hotsos2017cfp}) could inform evidence evaluation in incidence response because evaluating evidence properly is a primary scientific activity. 
While these \ac{DoD} projects ask the right questions about science to help with incident response, they have generally concluded security is not (yet) a science, and so there is little advice. 
\citet{spring2017why} argues this conclusion is ill-founded and excessively pessimistic.
However, the relevant point for this review is that the science of security literature does not advise \ac{CSIR}.

While the main part of the \ac{DoD} is out of scope, the \acl{IC} aspects of the \ac{US} federal government do provide adequate documents.
The \ac{DoD} and \ac{CIA} are generally not forthcoming with more conventional descriptions of their incident response practice. 
However, given that \ac{NIST} is not authoritative over the \ac{IC}, one would expect them to develop their own standard practices. 
Documents related to the practice of the \ac{US} \ac{IC} are occasionally published, with \ac{IC} attribution either explicit or implicit. 
Three such documents are relevant to evidence collection and analysis in incident investigation, forming what is essentially a \emph{de facto} standard. 
The first is a textbook published by the \ac{CIA} and used to train intelligence analysts \citep{heuer1999psychology} whose methods are applicable to \ac{CSIR}. 
The second is a pair of documents, the kill chain model of computer network attacks \citep{hutchins2011intelligence} and the diamond model of intrusion analysis \citep{caltagirone2013diamond}.
Unlike the textbook, these documents are not explicitly acknowledged as standard analysis methods within the defence and intelligence communities.  
However, the diamond model paper is published by the \ac{DoD} publisher, the Defense Technical Information Center.\footnote{See \url{http://www.dtic.mil/docs/citations/ADA586960}.} 
The diamond model builds on the kill chain. 
Given that Lockheed Martin, a US defence contractor, published the kill chain, it seems the papers are from overlapping communities.
Although it is tenuous to term three documents a `standard,' it is clear from the content that they come from a practitioner community and are one of the clearest available expressions of intrusion investigation methods publicly available. 
Therefore, it is necessary to place them in scope for discussion. 

The \ac{US} intelligence agencies exercise out-sized international influence.  The \ac{US} is part of an intelligence sharing alliance known as the five eyes, which includes Australia, Canada, New Zealand, and the United Kingdom. As the biggest partner in this alliance by far, what the \ac{US} intelligence practitioners do is probably accommodated, if not directly copied, by the other countries' services. 

US military influence goes beyond the five eyes. The \acf{NATO} is the biggest alliance the US leads, with 28 other countries. 
\ac{NATO} intelligence is also presumably influenced by five eyes, as Canada and the UK also play a big role. The US tends to supply logistics and intelligence support in its alliances, so intelligence standards are likely to influence allies. Other locations which cooperate extensively with the \ac{US} include Israel, South Korea, Japan, and the Philippines. 
By virtue of these alliances, it is reasonable to assume that intelligence professionals in all these places are relatively closely aligned with \ac{US} intelligence standards. 
These alliances end up including most of the global military and intelligence spending.  
Essentially only China and Russia are excluded, and the two of them account for 15-20\% of global military spending. 
Thus, although there are rather few \ac{IC} documents, and they are focused on the \ac{US}, they should provide information about how a large swath of such practitioners make decisions. 

In summary, this review will include the \ac{IETF}, \ac{ISO}, \ac{FIRST}, and available intelligence community documents as in-scope publication venues for incident investigation standards of practice for evidence collection, analysis, and reporting.  The review will exclude the \ac{ITU}, \ac{W3C}, \ac{ENISA} and US federal civilian government departments and agencies as out of scope due to either limited applicable content or limited jurisdiction. The most borderline organization is \ac{NIST}, which occasionally has standards canonized by the in-scope venues; the review will only include those \ac{NIST} standards cited or adopted explicitly by the four in-scope venues. 

Section~\ref{sub:Search-strategy} describes the method for determining which standards are relevant within these venues.

\section{Novelty\label{sec:Novelty}}

This section provides a brief structured survey of the survey literature
in order to demonstrate that our intended scope, as defined in Section~\ref{sec:Scope},
has not been previously surveyed. Lack of related surveys in two academic
venues provide evidence: IEEE Security and Privacy Systematization
of Knowledge (SoK) papers and ACM Computing Surveys (CSUR) journal.
All 35 extant SoK papers (as of August 1, 2017) are appraised for
relevance. For ACM CSUR we apply a keyword search to the corpus. 

IEEE S\&P has published 35 SoK papers since the venue initiated the
SoK publication option in 2010. We make an exhaustive evaluation of
relevance based on title and abstracts. Our basic relevance criterion
in this case is if the SoK is about designing or evaluating investigations
of maliciousness. Of the 35 SoK papers, only \citet{herley2017SoK}
and \citet{rossow2012prudent} are applicable to our project. I have
done prior work explaining why each of these is insufficient for our
purposes. \citet{spring2017why} addresses the shortcomings in \citet{herley2017SoK}.
\citet{hatleback2014exploring} expands and generalizes the good work
of \citet{rossow2012prudent}. None of the SoK papers systematize
knowledge of incident response, investigation, or analysis.

Our CSUR keyword search uses Google Scholar, limiting to publications
in ``ACM Computing Surveys'' between Jan 1, 2007 and Aug 1, 2017.
We use the keywords used in the main study, as described in Section~\ref{sub:Search-strategy}.
However, CSUR is a sufficiently different venue from our intended
scope that we sometimes find different keywords more useful. The surveys
returned by the following search terms are included in Table~\ref{tab:CSUR-results}.
Quotes are applied to the search as listed. 
\begin{enumerate}
\item ``computer security incident response''
\item ``incident investigation''
\item ``incident management''
\item ``computer security'' \& ``evidence collection''
\item ``incident response'' \& analysis
\item ``security incident'' \& investigation
\item ``computer security'' \& incident investigation 
\item ``computer security'' \& incident analysis
\end{enumerate}
\begin{table}[!th]
\begin{tabular}{|c|>{\centering}p{0.3cm}|>{\centering}p{0.3cm}|>{\centering}p{0.3cm}|>{\centering}p{0.3cm}|>{\centering}p{0.3cm}|>{\centering}p{0.3cm}|>{\centering}p{0.3cm}|>{\centering}p{0.3cm}||>{\centering}p{0.3cm}|>{\centering}p{0.3cm}|>{\centering}p{0.3cm}|}
\cline{2-12} 
\multicolumn{1}{c|}{} & \multicolumn{8}{c||}{Found in search \#} & \multicolumn{3}{c|}{Criteria}\tabularnewline
\hline 
Document & {\small{}1} & {\small{}2} & {\small{}3} & {\small{}4} & {\small{}5} & {\small{}6} & {\small{}7} & {\small{}8} & 1 & 2 & 3\tabularnewline
\hline 
\hline 
{\small{}\citet{Li:2017:DTC:computing}} &  &  & \cellcolor{darkgray} &  &  &  &  &  & \crossout{.1em}{.3cm} & \crossout{.1em}{.3cm} & \ding{51}\tabularnewline
\hline 
{\small{}\citet{Li:2017:DTD:disaster}} &  &  & \cellcolor{darkgray} &  &  &  &  &  & \ding{51} & \crossout{.1em}{.3cm} & \crossout{.1em}{.3cm}\tabularnewline
\hline 
{\small{}\citet{Jiang:2016:UGT:2911992.2906151}} &  &  &  & \cellcolor{darkgray} &  &  &  &  & \crossout{.1em}{.3cm} & \crossout{.1em}{.3cm} & \crossout{.1em}{.3cm}\tabularnewline
\hline 
{\small{}\citet{Jhaveri:2017:ARF:3022634.3003147}} &  &  &  &  & \cellcolor{darkgray} &  & \cellcolor{darkgray} & \cellcolor{darkgray} & \ding{51} & \crossout{.1em}{.3cm} & \ding{51}\tabularnewline
\hline 
{\small{}\citet{Pendleton:2016:SSS:3022634.3005714}} &  &  &  &  & \cellcolor{darkgray} &  & \cellcolor{darkgray} & \cellcolor{darkgray} & \crossout{.1em}{.3cm} & \crossout{.1em}{.3cm} & \crossout{.1em}{.3cm}\tabularnewline
\hline 
{\small{}\citet{Khan:2016:CLF:2911992.2906149}} &  &  &  &  & \cellcolor{darkgray} &  & \cellcolor{darkgray} & \cellcolor{darkgray} & \ding{51} & \ding{51} & \crossout{.1em}{.3cm}\tabularnewline
\hline 
{\small{}\citet{Laszka:2014:SII:2658850.2635673}} &  &  &  &  &  & \cellcolor{darkgray} & \cellcolor{darkgray} & \cellcolor{darkgray} & \crossout{.1em}{.3cm} & \ding{51} & \ding{51}\tabularnewline
\hline 
{\small{}\citet{Biddle:2012:GPL:2333112.2333114}} &  &  &  &  &  &  & \cellcolor{darkgray} & \cellcolor{darkgray} & \crossout{.1em}{.3cm} & \crossout{.1em}{.3cm} & \crossout{.1em}{.3cm}\tabularnewline
\hline 
{\small{}\citet{Milenkoski:2015:ECI:2808687.2808691}} &  &  &  &  &  &  & \cellcolor{darkgray} & \cellcolor{darkgray} & \crossout{.1em}{.3cm} & \ding{51} & \crossout{.1em}{.3cm}\tabularnewline
\hline 
{\small{}\citet{Tang:2016:ESP:2911992.2906153}} &  &  &  &  &  &  & \cellcolor{darkgray} &  & \crossout{.1em}{.3cm} & \crossout{.1em}{.3cm} & \crossout{.1em}{.3cm}\tabularnewline
\hline 
{\small{}\citet{Meng:2015:CSS:2808687.2785733}} &  &  &  &  &  &  & \cellcolor{darkgray} & \cellcolor{darkgray} & \crossout{.1em}{.3cm} & \crossout{.1em}{.3cm} & \crossout{.1em}{.3cm}\tabularnewline
\hline 
{\small{}\citet{Calzavara:2017:SWJ:3058791.3038923}} &  &  &  &  &  &  & \cellcolor{darkgray} & \cellcolor{darkgray} & \crossout{.1em}{.3cm} & \ding{51} & \crossout{.1em}{.3cm}\tabularnewline
\hline 
{\small{}\citet{Labati:2016:BRA:2966278.2933241}} &  &  &  &  &  &  & \cellcolor{darkgray} & \cellcolor{darkgray} & \crossout{.1em}{.3cm} & \ding{51} & \crossout{.1em}{.3cm}\tabularnewline
\hline 
{\small{}\citet{Ye:2017:SMD:3101309.3073559}} &  &  &  &  &  &  & \cellcolor{darkgray} & \cellcolor{darkgray} & \crossout{.1em}{.3cm} & \ding{51} & \crossout{.1em}{.3cm}\tabularnewline
\hline 
{\small{}\citet{Edwards:2015:SSO:2808687.2811403}{*}} &  &  &  &  &  &  & \cellcolor{darkgray} & \cellcolor{darkgray} & \ding{51} & \ding{51} & \ding{51}\tabularnewline
\hline 
{\small{}\citet{Avancha:2012:PMT:2379776.2379779}} &  &  &  &  &  &  & \cellcolor{darkgray} & \cellcolor{darkgray} & \crossout{.1em}{.3cm} & \crossout{.1em}{.3cm} & \crossout{.1em}{.3cm}\tabularnewline
\hline 
{\small{}\citet{Roy:2015:SCP:2737799.2693841}} &  &  &  &  &  &  & \cellcolor{darkgray} & \cellcolor{darkgray} & \crossout{.1em}{.3cm} & \ding{51} & \crossout{.1em}{.3cm}\tabularnewline
\hline 
{\small{}\citet{Chandola:2009:ADS:1541880.1541882}} &  &  &  &  &  &  & \cellcolor{darkgray} & \cellcolor{darkgray} & \ding{51} & \ding{51} & \crossout{.1em}{.3cm}\tabularnewline
\hline 
{\small{}\citet{Pearce:2013:VIS:2431211.2431216}} &  &  &  &  &  &  &  & \cellcolor{darkgray} & \ding{51} & \ding{51} & \crossout{.1em}{.3cm}\tabularnewline
\hline 
{\small{}\citet{Peng:2007:SND:1216370.1216373}} &  &  &  &  &  &  &  & \cellcolor{darkgray} & \crossout{.1em}{.3cm} & \ding{51} & \crossout{.1em}{.3cm}\tabularnewline
\hline 
{\small{}\citet{Younan:2012:RCC:2187671.2187679}} &  &  &  &  &  &  &  & \cellcolor{darkgray} & \crossout{.1em}{.3cm} & \ding{51} & \crossout{.1em}{.3cm}\tabularnewline
\hline 
{\small{}\citet{Egele:2008:SAD:2089125.2089126}{*}} &  &  &  &  &  &  &  & \cellcolor{darkgray} & \ding{51} & \ding{51} & \crossout{.1em}{.3cm}\tabularnewline
\hline 
\end{tabular}

\caption[Results of \acs*{ACM} CSUR search for extant literature reviews]{\label{tab:CSUR-results} Potentially relevant literature reviews from ACM CSUR. The three
relevance criteria are (1) relates to forensics rather than prediction; (2) technical, investigative focus; (3) useful level of abstraction of incident response. Papers with an asterisk ({*}) are discussed
in more detail in the text.}
\end{table}

These eight searches within CSUR return 22 unique results. Note in
particular that the two most precisely relevant searches return no
matches. As the search terms are expanded to include more general,
related terms, we find a handful of possibly relevant results. 

Two search terms were tried on CSUR but returned too many
clearly irrelevant results to be considered useful. Namely, \{``computer security'' \& analysis\} with 79 results and \{``computer security'' \& reporting\} with 25. Any relevant papers appear to be included in the
22 captured by Table~\ref{tab:CSUR-results}. 

To determine whether any of these 22 surveys already adequately cover
our topic of interest, we set out three relevance criteria. The survey
must:
\begin{enumerate}
\item relate to reconstructing past events (i.e., forensics) rather than
prediction;
\item focus on the technical- and knowledge-based decisions and processes,
rather than management processes;
\item use our target level of abstraction to discuss the problem of incident
response, investigation, or analysis (human decisions during the process),
rather than tool development, without being so abstract as to make
implementation impractical. 
\end{enumerate}
These criteria are marked in Table~\ref{tab:CSUR-results}, based
on each paper's abstract. Some papers deserved a look beyond their
abstracts.

\citet{Edwards:2015:SSO:2808687.2811403} is the only survey that
meets all three criteria, based on their abstract. However, their
focus is quite different from our intended focus. They discuss automation
of law-enforcement criminal investigation using computer science techniques.
There may be overlap with computer-security incident response, in
that some subset of law enforcement cases involve criminal action
against computers. However, the focus of \citet{Edwards:2015:SSO:2808687.2811403}
is what \citet[p.~3]{anderson2012measuring} call, quoting the European
Commission, ``traditional forms of crime... committed over electronic
communication networks and information systems.'' Incident response
and investigation focuses on a different category, ``crimes unique
to electronic networks,'' as well as organizational policy violations
that are not illegal under the relevant jurisdiction. Finally, \citet{Edwards:2015:SSO:2808687.2811403}
focus on automation of police tasks, whereas our focus would be on
the investigator's decision process in, among other things, choosing
which automation techniques to use and how to evaluate the evidence
they provide. These various differences make a clear case that our
intended survey topic is sufficiently distinct from \citet{Edwards:2015:SSO:2808687.2811403}. 

\citet[p.~1]{Egele:2008:SAD:2089125.2089126} aims to identify ``techniques
to assist human analysts in assessing ... whether a given sample deserves
closer manual inspection.'' It is, in fact, a survey of software
tools and their features, and does not discuss how an analyst should
use them. 

Many papers in Table~\ref{tab:CSUR-results} meet the technical criterion
(\#2) and fail the other two criteria. This pattern tends to be about
some specific subset of network defense -- for example, making better
passwords, intrusion detection systems, or web browser defenses. These
tools are certainly used and evaluated as part of security management,
and are important considerations. However, these details are tangential
to making decisions during incident response and investigation. 

These reviews of the available survey literature demonstrate a lacuna;
we lack a survey of incident response and investigation practices.
This omission matters. Most security research and security management
requires, directly or indirectly, ``ground truth'' evidence from
incident response teams. Research and management need this ``ground
truth'' evidence to evaluate any other security infrastructure, plans,
defenses, or policy. However, it seems there is no systematic review
of how this evidence should be collected, analyzed, and reported.
One must understand these steps in order to properly interpret any
such evidence. Therefore, although our topic is narrow, it has far-reaching
impact on information security more generally.

\section{Methods\label{sec:Methods}}

The scope of our topic is restricted to evidence collection, analysis, and reporting in human-driven computer security incident response.  We further restrict our review to internationally-recognized standards.  This choice keeps us as in touch as possible to actual professional practice without violating confidentiality around incident response, which organizations justifiably do not often disclose in detail.

We explain our methodology in three parts. First, our search strategy.
Secondly, our appraisal strategy for whether to include what we find in our synthesis. And finally, how we plan to synthesize the results into a coherent statement of current practice.

\subsection{Search strategy \label{sub:Search-strategy}}

The major determining factor in our literature search strategy is the scope of publication venues, as justified in Section~\ref{sec:Scope-Publication-Venues}.
Our search comes in two parts. First, we perform keyword searches in the relevant web archives. Secondly, we extract references from valid hits on these searches and include the referenced documents as sources to appraise.

The IC documents are selected by fiat, given the secretive nature of the community. Due to the idiosyncratic nature of the IC publication and publicity processes, there is no sense in a keyword search strategy.  We have arrived at the three core documents we shall consider as the ``standards'' from this community essentially by word of mouth through interaction with practitioners. 

Each of \ac{IETF},\footnote{\url{https://www.rfc-editor.org/search/rfc_search.php}} \ac{FIRST},\footnote{\url{https://first.org/standards/}} and \ac{ISO}\footnote{\url{https://www.iso.org/standards.html}} have dedicated web pages. For \ac{IETF} and \ac{ISO}, we use their site-based search engines that cover their respective corpora of standards. \ac{FIRST} has a smaller, more focused corpus of work, such that it does not have a site specific search engine; we use Google and prepend the term ``site:first.org'' to focus the search.

Quotes are applied to the search as listed. The keywords employed are: 
\begin{enumerate}
\item ``computer security incident response''
\item ``incident investigation''
\item ``incident management''
\item ``computer security'' \& ``evidence collection''
\item ``computer security'' \& analysis
\item ``computer security'' \& reporting
\end{enumerate}
We also added or modified terms slight to accommodate each search venue. The \ac{IETF} RFC search tool does not accommodate mixing quoted phrases with other terms, so for terms 4, 5, and 6 the quotes were removed. We added the following terms to the \ac{IETF} search: 
\begin{itemize}
\item ``incident response''
\end{itemize}
We added the following terms to the \ac{ISO} search, after it became apparent from searches 2 and 3 that the \ac{ISO} documents do not use the term ``computer security'' but rather ``information security'':
\begin{itemize}
\item ``information security'' \& ``evidence collection''
\item ``information security'' \& analysis
\item ``information security'' \& reporting
\end{itemize}
All the documents returned by this search strategy are appraised using the methods of Section~\ref{sub:Appraisal-strategy}. We then take a further search step and extract the references from those documents that pass the appraisal. We only include any cited documents which are publicly available (or, in the case of \ac{ISO}, readily available).
All the documents extracted from the references are appraised using the same methods to determine whether they are included in our review, independent of the document that cited it.

\subsection{Appraisal strategy\label{sub:Appraisal-strategy}}

The purpose of the appraisal is to determine whether each document is within the scope of evidence collection, analysis, and reporting for incident response. Our cutoff date for inclusion in the review is that the document must be published by July 1, 2017. 

Furthermore, because standards follow an orderly progression, they may be superseded or amended by future work. Drafts are also commonly published for public comment before being finalized. We exclude any standard superseded as of July 1, 2017, and incorporate any amendments finalized by July 1, 2017. We note the existence of drafts on new topics, but exclude their content from the review. 

Our specific inclusion criteria for whether the content is in-scope are the following:
\begin{itemize}
\item Target audience as expressed by author includes security professionals
\item Topic is within scope, namely it applies to one of the following parts
of computer-security incident response (or some clear synonym thereof)

\begin{itemize}
\item evidence collection
\item analysis
\item reporting
\end{itemize}
\item Topic is on investigator practice (rather than implementation of software or managerial considerations related to \acp{CSIRT})
\item Document is finalized (not a draft) and not explicitly superseded
as of July 1, 2017
\item Document is available in English
\end{itemize}
A document must satisfy all of these criteria in order to be included in the review.

\subsection{Synthesis methods\label{sub:Synthesis-methods}}

The input at this stage of the method will be all documents that are in scope; they will be documents for security professionals about investigator practice during evidence collection, analysis, and reporting.
Our synthesis goal is to evaluation the nature and quality of advice that these documents provide about making decisions during these phases of incident response.

As a prelude to our synthesis, we classify advice on these topics in several ways: to which phases the document applies, the directness with which the document applies to each phase, the type, investigative goals supported, broadness of scope, generalizability of advice, and formalism. We use this initial evaluation to identify groupings of documents and get an overview what the literature search has found to be available. The following describes each evaluation in more detail. 
\begin{description}
\item [{Phases}] indicates simply what combination of evidence \emph{collection},
\emph{analysis}, and \emph{reporting} the document covers. 
\item [{Directness}] has two possible values: \emph{direct} and \emph{constraints}.
Direct commentary on incident response explicitly talks about what an investigator should or should not do. Constraints provide only requirements for outcomes or outputs, and do not indicate how these properties should be achieved. Constraint-based advice is common when situating investigation within the larger context of incident response, and situating response within incident management. 
\item [{Type}] indicates what type and level of detail the document provides
to decision making. Possible values are \emph{case} \emph{study}, \emph{ontology}, \emph{advice}, and \emph{instructions}. At one end of the spectrum are case studies. Case studies report the facts of an individual case of investigation, without attempting to abstract up to general lessons. A categorization forms categories of useful actions (implicitly or explicitly from case studies), but gives no advice on how to apply these categories. Advice provides some ordering on what category of action should be taken, given certain conditions.  Finally, instructions provide explicit decision-making instructions on how to evaluate options. Type also provides some rough guide to how much effort it will take to apply the document to practice, with case studies being the most difficult. 
\item [{Goals}] indicates what sort of investigation the advice targets.
We distinguish three goals an investigator could have: \emph{fix} the system, gather \emph{intelligence} on adversaries, and make \emph{arrests}.
Certainly, there may be other goals, but these cover a wide degree of practical differences. Investigators need quite different information between these goals. For example, to fix a system, one needs to know everything that has been accessed by the adversary, but you need to know rather little about them. Whereas to make arrests, one cares very much about the adversary, but also is bound by several practical matters of what counts as admissible legal evidence of attribution and loss. When gathering intelligence on what an adversary may do next, these legal considerations fall away, but one also focuses on quite different aspects than fixing a system. For example, to gather adequate intelligence one need not enumerate all compromised systems. 
\item [{Scope}] reports how widely the document applies, as reported by the document. Options are \emph{narrow}, \emph{medium}, and \emph{broad}.
A narrowly-scoped document is intended to apply to only a small, non-representative group of people, and/or for a short period of time. Broad scopes are intended for most people within information security. Medium scope fits somewhere in between. Examples of medium scope are US-based law enforcement forensics specialists, or the operators of tier-three (that is, backbone) networks. 
\item [{Generalizability}] of advice indicates how likely the document can be relevant to contexts outside those for which it was specifically designed. Generalizability is explicitly level-set from the document's scope. Thus, a document with broad scope but no generalizability may still be applicable to more people than a narrowly-scoped document that is generalizable. Whereas scope is a measure taken directly from the document being evaluated, generalizability is an evaluation of potential not explicit in the document. Indicators of generalizability include use of models or methods from other disciplines with well-established other uses or evidence from sources other than the document itself that the advice from the document applies more widely. Options for this criterion are coarsely set as \emph{unlikely} $\left(<15\%,\pm5\%\right)$, \emph{likely} (in between unlikely and highly likely), and \emph{highly likely} $\left(>85\%,\pm5\%\right)$; values represent essentially the evaluator's prior belief on the document being applicable outside its stated scope. We have a final value, \emph{widely}, indicating the document is certainly generalizable beyond its scope and is likely to be generalizable to a much broader scope. 
\item [{Formalism}] reports the degree of mathematical formalism present in the advice provided. Options are \emph{none}, \emph{qualitative}, \emph{formal}, and \emph{perfunctory}. Perfunctory indicates formalization is present, but essentially unused to advance the arguments or positions of the document. This rating does not mean the formalism is wrong; however, it does indicate it would take significant effort on the part of the reader to make use of the formalism beyond what qualitative models would provide. None only applies to narratives that make no attempts at abstraction. Both qualitative models and formal mathematical models have value in their own ways, and one should not be considered preferred over the other per se. 
\end{description}

After this classification of the documents is complete, we undertake a more free form synthesis of the results. Our focus is on how investigators make decisions about evaluating the quality and importance of evidence, generalize from particular evidence to evidence of trends or patterns of behavior, and select what to report based on security constraints as well as what others will find most convincing. Although these align loosely with the three phases of investigation we have highlighted, we are not making a one-to-one connection between the three phases and our three focal points. For example, if an investigator knows some kind of evidence is particularly convincing to report, that should impact what they look for during the evidence collection phase. Section~\ref{sec:Discussion} reports the results of this classification, examination of focal points, and identification of preliminary gaps.

\subsection{Limitations\label{sub:Limitations}}

While these methods have much to recommend them, there are of course limitations. Some of these are practical, such as the restriction to publications available in English. Some limitations are a function of restricting the scope to standards. Perhaps the most dangerous limitation is a result of the subject matter -- security practitioners tend to be secretive about their methodologies. 

The restriction to English will naturally limit the results. For example, an internet search for 
(information security incident response) returns a couple dozen results on Google as well as Google Scholar. 
This seems to be the preferred term in mainland China, searching for computer security incident response 
returns only a couple of Taiwanese sites. 

The importance of this language choice on actually limiting documents available to us is less clear. EU and UN documents would be available in English as a matter of policy. The US government, which publishes in English dominates in this space, as do US companies. Countries that are not allied to the US and have developed computer security capabilities are relatively few; basically just the Russian Federation and the People's Republic of China (PRC). While it is possible these countries have published comprehensive decision making procedures for incident response, it seems unlikely. It also seems likely that, given how much attention the US security establishment pays to Russia and the PRC, if such a thing were published it would be found and reported on, if not translated. For these reasons, we judge the impact of limiting our search to English documents is a low risk. 

Focusing on standards, and what they cite, limits the scope but also creates other limitations. Indeed, reducing the scope to something manageable is one goal of focusing on standards, and this seems to function as intended. However, the type of information published in standards is different than that in academic journals and conferences, and this imposes some limits on the work. Specifically, standards are on a slower publication cycle than academic work. This delay would be a problem if our topic were covered much in the academic literature.
However, as indicated by Section~\ref{sec:Novelty}, academic publications do not appear to cover decision making during computer security incident response. 

Creation of technology standards is itself a complex process, and as Section~\ref{sec:Scope-Publication-Venues} touches on, the process has a complex history in its own right \citep{spring2011learned}.  The way standards are made imposes its own limitations on our findings.  Standards are rarely made purely for the dissemination of information; rather, they usually solidify a dominant business position. Security standards appear to be an outlier from this norm, as they have unique concerns about correctness and non-subversion~\citep{kuhlman2016trust}.  Incident response standards are essentially unstudied within the academic standards literature; \citet{kuhlman2016trust} mostly focus on cryptographic standards. This situation means we accept a risk in that we do not know what biases may be embedded in the creation of incident response standards. The standards literature provides evidence there will be a bias, but has not studied incident response standards in order to provide evidence for what that bias might be. We would like to know whose interests are best served by the creation of incident response standards, for example. 

A closely related limitation of concern involves secrecy. Many incident response organizations may not wish to disclose their processes and procedures in detail, lest the adversary learn how to subvert or avoid them. Other areas of information security experience similar publication restrictions. Incident responders likely have a legitimate concern in this regard, and may also have a legal or regulatory requirement to keep certain information or processes private. Therefore, this limitation imposes a significant risk that relevant information is not public. Lack of access obviously limits the review. Our approach to reduce the impact of this limitation is to understand that we ought to read between the lines of available documents when we can justify expanding our interpretation of a document's contents with circumstantial evidence from the context surrounding its publication. However, we must accept that there is an amount of information about our topic which simply is not public and we cannot hope to access for a public literature review. One could perhaps use news articles and audit reports to attempt to evidence the extent to which organizations in fact implement the available standards; we leave such investigations for future work.  We first need a baseline understanding of the standards literature from which to begin.

\section{Results \label{sec:Results}}

We present our results per search venue is Sections~\ref{sub:IETF-results} through~\ref{sub:IC-results}. Section~\ref{sub:results-Referenced-Documents} takes the results from these four venues, extracts the referenced documents from the results, and evaluates these cited documents. 

In summary, Table~\ref{tab:All-documents-relevant} lists the documents that are relevant to our review scope and purpose, and will analyze them in depth. We have selected these 29 documents from roughly 350 possible documents returned from search results and references. 
Many tables documenting our the decisions that lead to these results results can be found in the Appendix.

\begin{table}[!th]
\begin{tabular}{l|l}
RFC 2196, \S5.4 only \citep{rfc2196} & 27035-1:2016 \citet{iso27035-1-2016}\tabularnewline
\hline 
RFC 6545 \citep{rfc6545} & 27037 \citet{iso27037-2012}\tabularnewline
\hline 
RFC 7203 \citep{rfc7203} & 27041 \citet{iso27041-2015}\tabularnewline
\hline 
RFC 7970 \citep{rfc7970} & 27042 \citet{iso27042-2015}\tabularnewline
\hline 
RFC 8134 \citep{rfc8134} & 27043 \citet{iso27043-2015}\tabularnewline
\hline 
\ac{NIST} SP 800-61 r.2 \citep{cichonski2012_sp800-61} & \citet{kossakowski1999responding}\tabularnewline
\hline 
\ac{NIST} SP 800-86 \citep{nist800-86} & \citet{alberts2004defining}\tabularnewline
\hline 
\multicolumn{2}{l}{\ac{NIST} SP 800-83 r.1, \S4 only \citep{nist800-83r1}\emph{ }}\tabularnewline
\hline 
\citet{enisa2011proactive} & \citet{mundie2014incident}\tabularnewline
\hline 
\citet{hutchins2011intelligence} (The kill chain) & \citet{heuer1999psychology}\tabularnewline
\hline 
\citet{caltagirone2013diamond} (diamond model) & \citet{carrier2004event}\tabularnewline
\hline 
\citet{ciardhuain2004extended} & \citet[ch.~2]{casey2010handbook}\tabularnewline
\hline 
\citet{osorno2011coordinated} & \citet{leigland2004formalization}\tabularnewline
\hline 
\citet[ch.~5 only]{jp2-01-3} & \citet{stoll1988stalking}\tabularnewline
\hline 
\citet{cheswick1992evening} & \citet{mitropoulos2006incident}\tabularnewline
\end{tabular}

\caption{\label{tab:All-documents-relevant}All documents found to be relevant through the search methodology}
\end{table}

\subsection{IETF\label{sub:IETF-results}}

\begin{table}[t]
\begin{tabular}{|r|>{\centering}p{0.3cm}|>{\centering}p{0.3cm}|>{\centering}p{0.3cm}|>{\centering}p{0.3cm}|>{\centering}p{0.3cm}|}
\cline{2-6} 
\multicolumn{1}{r|}{} & \multicolumn{5}{c|}{Criteria}\tabularnewline
\hline 
Document & 1 & 2 & 3 & 4 & 5\tabularnewline
\hline 
\hline 
RFC 2350 \citep{rfc2350} & \ding{51} & \crossout{.1em}{.3cm} & \ding{51} & \ding{51} & \ding{51}\tabularnewline
\hline 
RFC 3607 \citep{rfc3607} & \ding{51} & \crossout{.1em}{.3cm} & \crossout{.1em}{.3cm} & \ding{51} & \ding{51}\tabularnewline
\hline 
RFC 5070 \citep{rfc5070} & \ding{51} & \ding{51} & \ding{51} & \crossout{.1em}{.3cm} & \ding{51}\tabularnewline
\hline 
RFC 6045 \citep{rfc6045} & \ding{51} & \ding{51} & \ding{51} & \crossout{.1em}{.3cm} & \ding{51}\tabularnewline
\hline 
RFC 6046 \citep{rfc6046} & \ding{51} & \ding{51} & \ding{51} & \crossout{.1em}{.3cm} & \ding{51}\tabularnewline
\hline 
RFC 6545 \citep{rfc6545} & \ding{51} & \ding{51} & \ding{51} & \ding{51} & \ding{51}\tabularnewline
\hline 
RFC 6546 \citep{rfc6546} & \ding{51} & \crossout{.1em}{.3cm} & \ding{51} & \ding{51} & \ding{51}\tabularnewline
\hline 
RFC 7203 \citep{rfc7203} & \ding{51} & \ding{51} & \ding{51} & \ding{51} & \ding{51}\tabularnewline
\hline 
RFC 7970 \citep{rfc7970} & \ding{51} & \ding{51} & \ding{51} & \ding{51} & \ding{51}\tabularnewline
\hline 
RFC 8134 \citep{rfc8134} & \ding{51} & \ding{51} & \ding{51} & \ding{51} & \ding{51}\tabularnewline
\hline 
\end{tabular}

\caption[\acs*{IETF} database search results]{\label{tab:IETF-results}\ac{IETF} database search results. The criteria
are (1) target audience is security professionals; (2) topic in scope,
per Section~\ref{sec:Scope-Topic}; (3) focus is investigator practices;
(4) document finalized and not obsoleted as of Aug 1, 2017; (5) available
in English. }
\end{table}

Table~\ref{tab:IETF-results} evaluates the results of our search
procedure. We pass four documents on, RFCs 6545, 7203, 7970, and 8134.

The \ac{IETF} documents break down into two clear broad categories, \ac{BCP}
21 on expectations for computer security incident
response \citep{rfc2350}, and all the others, which are to do with
\acf{IODEF}, its expansion, and usage.

As an expectations document, \ac{BCP} 21 focuses primarily on the services
and support a \ac{CSIRT} should provide to its constituency, who that constituency
should include, and so on. These considerations are vital to \ac{CSIRT}
operations; however they are not directly relevant to our question
at hand. 

The \ac{IODEF} projects in particular are feature \ac{CERT/CC} staff heavily,
Danyliw and Inacio during all their RFC authorship, and Trammell contributed
heavily to \acs{SiLK} (\acl{SiLK}) while at \ac{CERT/CC} before moving on. Because \ac{CERT/CC}
is also heavily involved in \ac{FIRST}, it is unsurprising to see a sort
of division of labor between the \ac{IETF} documents and the \ac{FIRST} documents.
In particular, the \ac{IODEF} format focuses almost exclusively on technical
issues of data exchange and reporting format. The softer considerations,
of how to collect, evaluate, and analyze the data contained within
\ac{IODEF} remain in the purview of \ac{FIRST}.

Thus, as technical reporting formats are in our scope of reporting
results, all RFCs related to \ac{IODEF} are relevant. There do not appear
to be any other \ac{IETF} documents within our scope. 

These \ac{IODEF} documents may at first seem to be out of scope, as we
specified that our scope is how investigators make decisions, not
what tools or formats they use to document them. This topic recurs
in Section~\ref{sub:IETF-referenced-docs}. \ac{IODEF} is essentially
a language for talking about computer security incidents. However,
because we are not interested in data formats, we are not interested
in the language per se. \ac{IODEF} is in scope because as a constructed
language it makes judgments about what aspects of incidents are important,
necessary, or possible to communicate. These judgments, at least implicitly,
bear on what an investigator should choose to report. We therefore
judge \ac{IODEF} as in-scope. However, we stress that our intended scope
is how to decide what information to report, not what language in
which to report it. Therefore, data formats and languages for anything
else remain out of scope.

\subsection{ISO\label{sub:ISO-results}}

Nine search terms return 31 total results, with 23 unique results
displayed in Table~\ref{tab:ISO-results}. ISO 27035-1, 27041, and 27043 meet our criteria to carry through to the citation-harvesting and synthesis stage: 
\begin{itemize}
\item \emph{Information security incident management, Part 1: Principles of incident management}  \citep{iso27035-1-2016}
\item \emph{Guidance on assuring suitability and adequacy of incident investigative
method} \citep{iso27041-2015}
\item \emph{Incident investigation principles and processes} \citep{iso27043-2015}
\end{itemize}

\begin{table}[th]
\begin{tabular}{|r|>{\centering}p{0.3cm}|>{\centering}p{0.3cm}|>{\centering}p{0.3cm}|>{\centering}p{0.3cm}|>{\centering}p{0.3cm}|}
\cline{2-6} 
\multicolumn{1}{r|}{} & \multicolumn{5}{c|}{Criteria}\tabularnewline
\hline 
Document & 1 & 2 & 3 & 4 & 5\tabularnewline
\hline 
\hline 
IEC 31010:2009  & \ding{51} & \crossout{.1em}{.3cm} & \ding{51} & \ding{51} & \ding{51}\tabularnewline
\hline 
\ac{ISO} 13485:2003 & \crossout{.1em}{.3cm} & \crossout{.1em}{.3cm} & \crossout{.1em}{.3cm} & \crossout{.1em}{.3cm} & \ding{51}\tabularnewline
\hline 
\ac{ISO}/IEC 17799:2005 & \ding{51} & \crossout{.1em}{.3cm} & \crossout{.1em}{.3cm} & \crossout{.1em}{.3cm} & \ding{51}\tabularnewline
\hline 
\ac{ISO}/IEC 27000:2009 & \ding{51} & \crossout{.1em}{.3cm} & \crossout{.1em}{.3cm} & \crossout{.1em}{.3cm} & \ding{51}\tabularnewline
\hline 
\ac{ISO}/IEC 27001:2005 & \ding{51} & \crossout{.1em}{.3cm} & \crossout{.1em}{.3cm} & \crossout{.1em}{.3cm} & \ding{51}\tabularnewline
\hline 
\ac{ISO}/IEC 27002:2005 & \ding{51} & \crossout{.1em}{.3cm} & \crossout{.1em}{.3cm} & \crossout{.1em}{.3cm} & \ding{51}\tabularnewline
\hline 
\ac{ISO}/IEC 27004:2016 & \ding{51} & \crossout{.1em}{.3cm} & \crossout{.1em}{.3cm} & \ding{51} & \ding{51}\tabularnewline
\hline 
\ac{ISO}/IEC 27005:2011 & \ding{51} & \crossout{.1em}{.3cm} & \crossout{.1em}{.3cm} & \ding{51} & \ding{51}\tabularnewline
\hline 
\ac{ISO}/IEC 27006:2011 & \crossout{.1em}{.3cm} & \crossout{.1em}{.3cm} & \crossout{.1em}{.3cm} & \crossout{.1em}{.3cm} & \ding{51}\tabularnewline
\hline 
\ac{ISO}/IEC 27006:2015 & \crossout{.1em}{.3cm} & \crossout{.1em}{.3cm} & \crossout{.1em}{.3cm} & \ding{51} & \ding{51}\tabularnewline
\hline 
\ac{ISO}/IEC 27033-1:2009 & \ding{51} & \crossout{.1em}{.3cm} & \crossout{.1em}{.3cm} & \crossout{.1em}{.3cm} & \ding{51}\tabularnewline
\hline 
\ac{ISO}/IEC 27033-4:2014 & \ding{51} & \crossout{.1em}{.3cm} & \crossout{.1em}{.3cm} & \ding{51} & \ding{51}\tabularnewline
\hline 
\ac{ISO}/IEC 27035:2011 & \ding{51} & \ding{51} & \ding{51} & \crossout{.1em}{.3cm} & \ding{51}\tabularnewline
\hline 
\ac{ISO}/IEC 27035-1:2016 & \ding{51} & \ding{51} & \ding{51} & \ding{51} & \ding{51}\tabularnewline
\hline 
\ac{ISO}/IEC 27035-2:2016 & \ding{51} & \crossout{.1em}{.3cm} & \crossout{.1em}{.3cm} & \ding{51} & \ding{51}\tabularnewline
\hline 
\ac{ISO}/IEC 27041:2015 & \ding{51} & \ding{51} & \ding{51} & \ding{51} & \ding{51}\tabularnewline
\hline 
\ac{ISO}/IEC 27043:2015 & \ding{51} & \ding{51} & \ding{51} & \ding{51} & \ding{51}\tabularnewline
\hline 
\ac{ISO}/IEC TR 18044:2004 & \crossout{.1em}{.3cm} & \ding{51} & \ding{51} & \crossout{.1em}{.3cm} & \ding{51}\tabularnewline
\hline 
\ac{ISO}/IEC TR 20004:2015 & \ding{51} & \crossout{.1em}{.3cm} & \crossout{.1em}{.3cm} & \ding{51} & \ding{51}\tabularnewline
\hline 
\ac{ISO}/NP TS 11633-1 & \ding{51} & \crossout{.1em}{.3cm} & \crossout{.1em}{.3cm} & \crossout{.1em}{.3cm} & \ding{51}\tabularnewline
\hline 
\ac{ISO}/TR 11633-1:2009 & \ding{51} & \crossout{.1em}{.3cm} & \crossout{.1em}{.3cm} & \ding{51} & \ding{51}\tabularnewline
\hline 
\ac{ISO}/TR 11633-2:2009 & \ding{51} & \crossout{.1em}{.3cm} & \crossout{.1em}{.3cm} & \ding{51} & \ding{51}\tabularnewline
\hline 
\ac{ISO}/TS 19299:2015 & \crossout{.1em}{.3cm} & \crossout{.1em}{.3cm} & \crossout{.1em}{.3cm} & \ding{51} & \ding{51}\tabularnewline
\hline 
\end{tabular}

\caption[\acs*{ISO} database search results]{\label{tab:ISO-results}ISO database search results. 
The criteria are (1) target audience is security professionals; (2) topic in scope,
per Section~\ref{sec:Scope-Topic}; (3) focus is investigator practices;
(4) document finalized and not obsoleted as of Aug 1, 2017; (5) available
in English. }
\end{table}

\subsection{FIRST\label{sub:FIRST-results}}

\ac{FIRST} is the smallest body surveyed, and it is not primarily a standards
organization but rather a forum for organizations with a shared purpose
-- incident response.

\begin{table}[t]
\begin{tabular}{|r|>{\centering}p{0.3cm}|>{\centering}p{0.3cm}|>{\centering}p{0.3cm}|>{\centering}p{0.3cm}|>{\centering}p{0.3cm}|}
\cline{2-6} 
\multicolumn{1}{r|}{} & \multicolumn{5}{c|}{Criteria}\tabularnewline
\hline 
Document & 1 & 2 & 3 & 4 & 5\tabularnewline
\hline 
\hline 
\citet{mundie2014incident} & \ding{51} & \ding{51} & \ding{51} & \ding{51} & \ding{51}\tabularnewline
\hline 
\citet{alberts2004defining} & \ding{51} & \ding{51} & \ding{51} & \ding{51} & \ding{51}\tabularnewline
\hline 
OCTAVE \citep{Caralli2007octave} & \ding{51} & \crossout{.1em}{.3cm} & \crossout{.1em}{.3cm} & \ding{51} & \ding{51}\tabularnewline
\hline 
\citet{enisa2006guide} & \crossout{.1em}{.3cm} & \ding{51} & \ding{51} & \ding{51} & \ding{51}\tabularnewline
\hline 
\citet{janet_sysadmin} & \ding{51} & \crossout{.1em}{.3cm} & \crossout{.1em}{.3cm} & \ding{51} & \ding{51}\tabularnewline
\hline 
\citet{enisa2011proactive} & \ding{51} & \ding{51} & \ding{51} & \ding{51} & \ding{51}\tabularnewline
\hline 
\citet{cichonski2012_sp800-61} & \ding{51} & \ding{51} & \ding{51} & \ding{51} & \ding{51}\tabularnewline
\hline 
\citet{etsi2014indicators} & \ding{51} & \crossout{.1em}{.3cm} & \ding{51} & \ding{51} & \ding{51}\tabularnewline
\hline 
RFC 2350 \citep{rfc2350} & \ding{51} & \crossout{.1em}{.3cm} & \ding{51} & \ding{51} & \ding{51}\tabularnewline
\hline 
RFC 2196 \citep[\S5.4 only]{rfc2196} & \ding{51} & \ding{51} & \ding{51} & \ding{51} & \ding{51}\tabularnewline
\hline 
RFC 2827 \citep{rfc2827} & \ding{51} & \crossout{.1em}{.3cm} & \crossout{.1em}{.3cm} & \ding{51} & \ding{51}\tabularnewline
\hline 
RFC 2504 \citep{rfc2504} & \crossout{.1em}{.3cm} & \ding{51} & \ding{51} & \ding{51} & \ding{51}\tabularnewline
\hline 
\end{tabular}

\caption[Summary of results found through \acs*{FIRST}]{\label{tab:FIRST-results}\ac{FIRST} results summary. The criteria are
(1) target audience is security professionals; (2) topic in scope,
per Section~\ref{sec:Scope-Topic}; (3) focus is investigator practices;
(4) document finalized as of Aug 1, 2017; (5) available in English. }
\end{table}

On it's ``standards'' web page, \ac{FIRST} lists four standards it maintains:
\begin{itemize}
\item \acf{TLP} on agreed-upon levels for marking
information sensitivity
\item \acf{CVSS} on describing the
characteristics and severity of defects in software systems (not to be confused with \ac{CWSS} by \ac{MITRE})
\item \acf{IEP} is a reporting format; in this
regard it is another language for reporting, similar to \ac{IODEF} or those listed in Table~\ref{tab:IETF-related-cite-acronyms}. \ac{IEP}'s focus
is on disseminating information responsibly and quickly during incident response. %
\item \acf{pDNS} is a formatting standard
for \acs{DNS} traffic analysis; the \ac{FIRST} group is working on an \ac{IETF} standard. 
\end{itemize}
None of these standards meet our relevance criteria, because none are about investigator practice.
They are all things a competent investigator
should know how to interact with and interpret, but they do not help
us understand what decisions an investigator should make in a given
scenario. \ac{FIRST} also notes it contributes to several \ac{ISO} standards,
which Section~\ref{sub:ISO-results} covers (namely, 27010, 27032,
27035, 27037, and 29147). 

More instructive than these standards are \ac{FIRST}'s ``Security Reference
Index'' that are ``helpful'' to the \ac{FIRST} community.\footnote{\url{https://first.org/resources/guides/reference}}
\ac{FIRST}'s members are many, if not most, of the professionals and practitioners
that we hope to come to understand. The documents Table~\ref{tab:FIRST-results}
evaluates are listed as either best practices or standards in this
reference index.\footnote{Strictly speaking, \citet{mundie2014incident} and \citet{alberts2004defining}
are not linked directly; they are the most relevant part of a suite
of publications linked to by \ac{FIRST} as \url{https://www.cert.org/incident-management/publications/index.cfm}.} Five documents emerge as relevant to our review: \citet{alberts2004defining,cichonski2012_sp800-61,rfc2196,enisa2011proactive,mundie2014incident}. 

The security reference index also links to the home pages of other
security organizations; however, we cannot review the contents of
everything these organizations have produced in full. In large part,
the information is more about solving specific technical problems
than our target for a general problem solving method. Such specific
problems make for instructive cases when thinking about generalized
methods, and so these organizations do provide an integral function
to our topic. But they do not aim for the types of documents in scope
of our review. The organizations identified are:
\begin{itemize}
\item \acf{CAIDA}, \url{www.caida.org}
\item \ac{CERT/CC} (formerly Computer Emergency Response Team), \url{www.cert.org}
\item \acf{CIS} Benchmarking, \url{http://www.cisecurity.org/}
\item Team Cymru, a security think tank, \url{https://www.team-cymru.org/services.html}
\item \acf{ENISA},
\url{https://www.enisa.europa.eu/}, including \ac{CSIRT} services
\url{https://www.enisa.europa.eu/topics/csirt-cert-services}
\item \ac{OWASP}, \url{https://www.owasp.org/index.php/OWASP_Guide_Project}
\item Microsoft Security Guidance Center \url{https://technet.microsoft.com/en-us/library/cc184906.aspx}
\item \acf{SANS} reading room, \url{https://www.sans.org/reading-room/}
\end{itemize}
Although \citet{enisa2006guide} targets management rather than practitioners,
it provides links to training for practitioners. The two organizations
the report lists are \ac{CERT/CC} and the EU-funded TRANSITS.

\subsection{Intelligence community\label{sub:IC-results}}

The canonical training course for \ac{CIA} and other intelligence analysts
is \citet{heuer1999psychology}. The book is essentially applied psychology.
It covers topics such as analyzing competing hypotheses, which includes
evaluating whether evidence has been planted to deceive, as well as
overcoming human cognitive biases such as anchoring, vividness, and
oversensitivity to consistency. Such methods, especially for evaluating
evidence in the face of deception, have clear relevance to incident
investigation. 

The model of a computer attack as following a predictable `kill chain'
of steps from start to finish was published by Lockheed Martin incident
responders \citep{hutchins2011intelligence}. The seven steps are
reconnaissance, weaponization, delivery, exploitation, installation,
command-and-control, and actions on objectives. These are the steps
in one single attack -- a single phishing email, a single drive-by
download with a malicious advert, etc. Adversaries almost always compose
a campaign out of multiple attacks; the objectives of one attack may
be to obtain a platform from which further attacks are possible. The
purpose of this model ``is to capture something useful about the
pattern all, or at least nearly all, attacks follow,'' so the analyst
can anticipate what to look for or expect next \citep[p.~10]{spring2016thinking}. 

\citet{caltagirone2013diamond} builds on attack ontologies, specifically
the kill chain, and intelligence analysis to perform attribution in
computer security incidents and analysis of whole campaigns. The method
incorporates Bayesian statistics to model belief updates of the analyst.
These statistical details are explicitly intended to help overcome
analyst cognitive biases, such as those discussed in \citet{heuer1999psychology}.

\begin{table}[t]
\begin{tabular}{|r|>{\centering}p{0.3cm}|>{\centering}p{0.3cm}|>{\centering}p{0.3cm}|>{\centering}p{0.3cm}|>{\centering}p{0.3cm}|}
\cline{2-6} 
\multicolumn{1}{r|}{} & \multicolumn{5}{c|}{Criteria}\tabularnewline
\hline 
Document & 1 & 2 & 3 & 4 & 5\tabularnewline
\hline 
\hline 
\citet{heuer1999psychology} & \ding{51} & \ding{51} & \ding{51} & \ding{51} & \ding{51}\tabularnewline
\hline 
\citet{hutchins2011intelligence} & \ding{51} & \ding{51} & \ding{51} & \ding{51} & \ding{51}\tabularnewline
\hline 
\citet{caltagirone2013diamond} & \ding{51} & \ding{51} & \ding{51} & \ding{51} & \ding{51}\tabularnewline
\hline 
\end{tabular}

\caption[\acl*{IC} search results]{\label{tab:IC-results}Intelligence community results summary. The
criteria are (1) target audience is security professionals; (2) topic
in scope, per Section~\ref{sec:Scope-Topic}; (3) focus is investigator
practices; (4) document finalized as of Aug 1, 2017; (5) available
in English. }
\end{table}

\FloatBarrier

\subsection{Referenced Documents\label{sub:results-Referenced-Documents}}

We harvest citations from the standards identified as relevant in
the prior parts of Section~\ref{sec:Results}. We summarize the results
here; the evaluations are detailed in a subsection for each publication
venue. The documents harvested from citations that are directly relevant
to our review are:
\begin{multicols}{2}
\begin{itemize}
\item \citet{carrier2004event}
\item \citet[ch.~2]{casey2010handbook}
\item \citet{ciardhuain2004extended}
\item \citet{leigland2004formalization}
\item 27037 \citet{iso27037-2012}
\item 27042 \citet{iso27042-2015}
\item \ac{NIST} SP 800-83 rev 1, \S4 only \citep{nist800-83r1}\emph{ }
\item \ac{NIST} SP 800-86 \citep{nist800-86}
\item \citet{osorno2011coordinated}
\item \citet{kossakowski1999responding}
\item \citet{cheswick1992evening}
\item \citet{stoll1988stalking}
\item \citet{mitropoulos2006incident}
\item \citet[ch.~5 only]{jp2-01-3}
\end{itemize}
\end{multicols}

\subsubsection{IETF documents\label{sub:IETF-referenced-docs}}

The four relevant \ac{IETF} standards reference 97 unique documents, excluding
the \ac{IODEF}-related standards already considered in Section~\ref{sub:IETF-results}.
These documents fall into three broad categories: technical implementation
requirements and dependencies; other related computer-security-incident
report formats; and broader incident handling guidelines that describe
the larger analysis and response context within which the reporting
formats are used. This first category of implementation dependencies
is not relevant to our project. Therefore, we focus on other reporting
documents and broader incident handling guidelines. There are 22 cited
reporting-related documents and exactly one related to broader incident
handling and use of the reporting formats.

Table~\ref{tab:IETF-related-cite-acronyms} lists the reporting formats
and data sources cited by the \ac{IETF} results in Section~\ref{sub:IETF-results}.
Other report formats are primarily produced by \ac{NIST} and MITRE -- with
funding from the US government including \ac{NIST}. These projects also
include the only referenced documents that are continuously updated
data archives. The data formats for \ac{CVE}, \ac{CWE}, \ac{CWSS}, \ac{CCE}, \ac{CCSS}, \ac{CPE}
are also continuously populated and published by \ac{NIST} and MITRE as
new vulnerabilities, platforms, etc. are discovered or developed.
Thus these projects provide not only a format, but a standard reference
dictionary of the possible terms with which the format may be populated.
CVSS is perhaps the most important of these metrics which provide
data and scoring; for a survey that relates \ac{CVSS} to other security
metrics, see \citet{Pendleton:2016:SSS:3022634.3005714}.

These standard dictionaries are referenced by many of the other data
formats which inherit the field essentially as a data type. For example,
\ac{MAEC} may indicate which vulnerability a malware targets using its
\ac{CVE} number. Such dictionaries are useful background knowledge during
incident response, and help reduce confusion by providing common
reference tags. However, following the pattern of other documents
surveyed, these reference dictionaries do not provide agreed-upon
evidence collection, field population, or analysis guidelines for
their contents. 

The next largest group of cited work from identified RFCs are three
more \ac{IETF} documents related to \ac{IODEF} that did not appear in the original
search. Other documents are also related to \ac{IODEF}. CVRF and XEP-0268
extend and implement \ac{IODEF}, respectively. AirCERT is a proposed implementation
architecture that uses \ac{IODEF} in automated indicator exchange. Our
conclusions about the \ac{IODEF} results identified in Section~\ref{sub:IETF-results}
therefore apply equally to these other documents, and we do not need
to pay them special attention. 

The remaining documents fall loosely into the \ac{NIST}-MITRE orbit. \ac{ISO}
19770 for asset management is developed separately from, but is related
to, CPE and CCE, \ac{NIST}'s asset management for platforms and configurations,
respectively. MMDEF is not directly related to MAEC; however, MAEC
has adopted a significant component of the MMDEF schema. 

The only citation related to actual decision making during an incident
is \ac{NIST} SP 800-61 \citep{cichonski2012_sp800-61}. This document is
already included from the \ac{FIRST} results, see Section~\ref{sub:FIRST-results}.
Thus, from the \ac{IETF} citations, we add no new documents to the review.

\begin{table}[th]
\begin{tabular}{|c|>{\centering}p{1.95cm}|>{\RaggedRight}p{7cm}|}
\hline 
\textbf{Publisher} & \textbf{Type} & \textbf{Name}\tabularnewline
\hline 
\hline 
\acs{CERT/CC} & Architecture & \hskip 1mm minus 1mm \acf{AirCERT}\tabularnewline
\hline 
ICASI & Format & \hskip 1mm minus 1mm \acf{CVRF}\tabularnewline
\hline 
\acs{IEEE} & Format & \hskip 1mm minus 1mm \acf{MMDEF}\tabularnewline
\hline 
\multirow{3}{*}{\ac{IETF}} & Format & \hskip 1mm minus 1mm \acf{IDMEF}\tabularnewline
\cline{2-3} 
 & Format & \hskip 1mm minus 1mm \acf{iodefplus}\tabularnewline
\cline{2-3} 
 & Format & RFC 5941, Sharing Transaction Fraud Data (extends \ac{IODEF})\tabularnewline
\hline 
\ac{ISO} & Format & Software asset management: Software identification tag (ISO 19770)\tabularnewline
\hline 
\ac{FIRST} & Data & \hskip 1mm minus 1mm \acf{CVSS}\tabularnewline
\hline 
\multirow{7}{*}{\acs{MITRE}} & Format & \hskip 1mm minus 1mm \acf{CAPEC}\tabularnewline
\cline{2-3} 
 & Format & \hskip 1mm minus 1mm \acf{CEE}\tabularnewline
\cline{2-3} 
 & Data & \hskip 1mm minus 1mm \acf{CVE}\tabularnewline
\cline{2-3} 
 & Data & \hskip 1mm minus 1mm \acf{CWE}\tabularnewline
\cline{2-3} 
 & Data & \hskip 1mm minus 1mm \acf{CWSS}\tabularnewline
\cline{2-3} 
 & Format & \hskip 1mm minus 1mm \acf{MAEC}\tabularnewline
\cline{2-3} 
 & Format & \hskip 1mm minus 1mm \acf{OVAL}\tabularnewline
\hline 
\multirow{6}{*}{\ac{NIST}} & Data & \hskip 1mm minus 1mm \acf{CCE}\tabularnewline
\cline{2-3} 
 & Data & \hskip 1mm minus 1mm \acf{CCSS} \tabularnewline
\cline{2-3} 
 & Data & \hskip 1mm minus 1mm \acf{CPE}\tabularnewline
\cline{2-3} 
 & Format & \hskip 1mm minus 1mm \acf{OCIL}\tabularnewline
\cline{2-3} 
 & Format & \hskip 1mm minus 1mm \acf{SCAP} \tabularnewline
\cline{2-3} 
 & Format & \hskip 1mm minus 1mm \acf{XCCDF}\tabularnewline
\hline 
XMPP & Format & XEP-0268 Incident Handling (using \ac{IODEF})\tabularnewline
\hline 
\end{tabular}

\caption[Data formats cited by \acs*{IETF} standards]{\label{tab:IETF-related-cite-acronyms}Computer-security related reporting
formats and data formats cited by \acs*{IETF} standards documents. }
\end{table}

\subsubsection{ISO documents\label{sub:ISO-referenced}}

We extract the references from the three relevant \ac{ISO} standards, namely
\ac{ISO}/IEC 27035, \ac{ISO}/IEC 27041:2015, \ac{ISO}/IEC 27043:2015. \ac{ISO}/IEC 27035
comes in two parts; although only the first part is in scope, we extract
references from both parts. Although \ac{ISO} charges for access to its
documents, all the bibliographies are freely available, so we include
all documents in this step. 

There are 81 total references among the documents, with 64 unique
references. Of these, 20 are elements of the \ac{ISO}/IEC 27000 series
of standards explicitly targeting information security. Four are the
documents already referenced, and several others are already noted
as not relevant in Table~\ref{tab:ISO-results}. However, from the
references we add the following two 27000-series publications to our survey documents as relevant to our survey (27037 and 27042):
\begin{itemize}
\item \emph{Guidelines for identification, collection, acquisition and preservation of digital evidence} \citep{iso27037-2012}
\item \emph{Guidelines for the analysis and interpretation of digital evidence} \citep{iso27042-2015}
\end{itemize}
Of the remaining 44 referenced documents, 13 are further \ac{ISO} standards.
Specifically: 

\begin{minipage}[t]{1\columnwidth}%
\begin{multicols}{2}
\begin{itemize}
\item[] \ac{ISO} 15489-1
\item[] \ac{ISO} 8601
\item[] \ac{ISO} 9000
\item[] \ac{ISO}/IEC 10118-2
\item[] \ac{ISO}/IEC 12207:2008 
\item[] \ac{ISO}/IEC 17024:2012
\item[] \ac{ISO}/IEC 17025:2005
\item[] \ac{ISO}/IEC 17043:2010
\item[] \ac{ISO}/IEC 20000
\item[] \ac{ISO}/IEC 29147
\item[] \ac{ISO}/IEC 30111
\item[] \ac{ISO}/IEC 30121
\item[] \ac{ISO}/IEC/IEEE 29148:2011
\end{itemize}
\end{multicols}%
\end{minipage}
\\

All of these other \ac{ISO} documents are out of scope. We can further
remove the following as unavailable or already evaluated: ILAC-G19,
which directly follows from \ac{ISO} 17020 and 17025; RFC 5070 \citep{rfc5070},
see Section~\ref{sub:IETF-results}; one by Valjarevic and Venter
that is not available but appears by title and timing to be a working
group presentation discussing the other two papers by these authors.
We also will not consider the Daubert 1993 US Supreme Court case,
as we aim to be jurisdiction neutral. Removing these leaves the documents
listed and evaluated for relevance in Table~\ref{tab:ISO-referenced}.
We pass the following documents on to the next stage of analysis:
\citet{carrier2004event,casey2010handbook,ciardhuain2004extended,leigland2004formalization}.
And, like the \ac{IETF}, \ac{ISO} cites \ac{NIST} SP 800-61 \citep{cichonski2012_sp800-61}.

\begin{table}[th]
\begin{tabular}{|r|>{\centering}p{0.3cm}|>{\centering}p{0.3cm}|>{\centering}p{0.3cm}|>{\centering}p{0.3cm}|>{\centering}p{0.3cm}|}
\cline{2-6} 
\multicolumn{1}{r|}{} & \multicolumn{5}{c|}{Criteria}\tabularnewline
\hline 
Document & 1 & 2 & 3 & 4 & 5\tabularnewline
\hline 
\hline 
\citet{ACPO2012} & \crossout{.1em}{.3cm} & \ding{51} & \ding{51} & \ding{51} & \ding{51}\tabularnewline
\hline 
\citet{valjarevicanalyses} & \ding{51} & \ding{51} & \ding{51} & \crossout{.1em}{.3cm} & \ding{51}\tabularnewline
\hline 
\citet{valjarevic2012harmonised} & \ding{51} & \ding{51} & \ding{51} & \crossout{.1em}{.3cm} & \ding{51}\tabularnewline
\hline 
\citet{carrier2003getting} & \ding{51} & \ding{51} & \ding{51} & \crossout{.1em}{.3cm} & \ding{51}\tabularnewline
\hline 
\citet{carrier2004event} & \ding{51} & \ding{51} & \ding{51} & \ding{51} & \ding{51}\tabularnewline
\hline 
\citet{national2009strengthening} & \crossout{.1em}{.3cm} & \crossout{.1em}{.3cm} & \crossout{.1em}{.3cm} & \ding{51} & \ding{51}\tabularnewline
\hline 
\citet{casey2010handbook} & \ding{51} & \ding{51} & \ding{51} & \ding{51} & \ding{51}\tabularnewline
\hline 
\citet{cohen2010fundamentals} & \crossout{.1em}{.3cm} & \crossout{.1em}{.3cm} & \ding{51} & \ding{51} & \ding{51}\tabularnewline
\hline 
\citet{cohen2011state} & \crossout{.1em}{.3cm} & \crossout{.1em}{.3cm} & \ding{51} & \ding{51} & \ding{51}\tabularnewline
\hline 
\citet{palmer2001road} & \crossout{.1em}{.3cm} & \ding{51} & \crossout{.1em}{.3cm} & \crossout{.1em}{.3cm} & \ding{51}\tabularnewline
\hline 
\citet{pollitt2008applying} & \crossout{.1em}{.3cm} & \ding{51} & \crossout{.1em}{.3cm} & \ding{51} & \ding{51}\tabularnewline
\hline 
\citet{reith2002examination} & \crossout{.1em}{.3cm} & \ding{51} & \crossout{.1em}{.3cm} & \ding{51} & \ding{51}\tabularnewline
\hline 
\citet{beebe2005hierarchical} & \crossout{.1em}{.3cm} & \ding{51} & \ding{51} & \ding{51} & \ding{51}\tabularnewline
\hline 
\citet{ciardhuain2004extended} & \ding{51} & \ding{51} & \ding{51} & \ding{51} & \ding{51}\tabularnewline
\hline 
\citet{leigland2004formalization} & \ding{51} & \ding{51} & \ding{51} & \ding{51} & \ding{51}\tabularnewline
\hline 
\citet{rowlingson2004ten} & \ding{51} & \crossout{.1em}{.3cm} & \ding{51} & \ding{51} & \ding{51}\tabularnewline
\hline 
\citet{hankins2009comparative} & \crossout{.1em}{.3cm} & \crossout{.1em}{.3cm} & \ding{51} & \ding{51} & \ding{51}\tabularnewline
\hline 
\citet{swgde2009nasresponse} & \crossout{.1em}{.3cm} & \crossout{.1em}{.3cm} & \crossout{.1em}{.3cm} & \ding{51} & \ding{51}\tabularnewline
\hline 
\citet{garfinkel2009bringing} & \ding{51} & \crossout{.1em}{.3cm} & \crossout{.1em}{.3cm} & \ding{51} & \ding{51}\tabularnewline
\hline 
\citet{technical2001electronic} & \crossout{.1em}{.3cm} & \crossout{.1em}{.3cm} & \ding{51} & \ding{51} & \ding{51}\tabularnewline
\hline 
\citet{alberts2014Introduction} & \crossout{.1em}{.3cm} & \crossout{.1em}{.3cm} & \crossout{.1em}{.3cm} & \ding{51} & \ding{51}\tabularnewline
\hline 
\citet{cichonski2012_sp800-61} & \ding{51} & \ding{51} & \ding{51} & \ding{51} & \ding{51}\tabularnewline
\hline 
\end{tabular}

\caption[\acs*{ISO} database search results]{\label{tab:ISO-referenced}\acs*{ISO} database search results. The criteria
are (1) target audience is security professionals; (2) topic in scope,
per Section~\ref{sec:Scope-Topic}; (3) focus is investigator practices;
(4) document finalized and not obsoleted as of Aug 1, 2017; (5) available
in English.}
\end{table}

Also cited are two further MITRE data formats not covered in Section~\ref{sub:IETF-referenced-docs}:
STIX (Structured Threat Information eXpression) and TAXII (Trusted
Automated eXchange of Indicator Information). These build on MAEC,
CVE, and so on as formats for exchanging incident data. Like the other
reporting formats already discussed, STIX and TAXII are not directly
relevant to our incident response decision-making topic. They
are a language in which to do reporting; reporting is in scope. However,
we plan to discuss what to report in a language-independent way. 

The ACPO (Association of Chief Police Officers) guidelines are representative
of many of the documents in Table~\ref{tab:ISO-referenced}. Their
target audience is ``UK law enforcement personnel who may deal with
digital evidence'' \citep[p.~6]{ACPO2012}. It is primarily about
the legal chain of custody necessary to bring digital evidence to
court. This topic is about evidence collection, so it is in scope.
But the target audience is law enforcement, not security practitioners.
The guidelines are not transferable to our topic of interest. The
extent of comment on the actual work of understanding what the digital
evidence means is constrained to ``it is not practically possible
to examine every item of digital data and clear tasking is needed
to ensure that the digital forensic practitioner has the best chance
of finding any evidence which is relevant to the investigation''
\citep[p.~10]{ACPO2012}. 

Some relevant work will not be carried through because it is obsoleted
in a rather round-about way. \citet[p.~1]{valjarevic2012harmonised}
notes ``an effort to \foreignlanguage{british}{standardise} the process
has started within \ac{ISO}, by the authors.''
Thus we consider papers by these authors to be obsolete because the
authors directly subsumed their ideas into the \ac{ISO} process. The process
classes and activities used by \ac{ISO} are clearly derived from \citet[p.~6]{valjarevicanalyses},
which also contains a matrix of how these reference terms relate to
other common forensic investigation ontologies. This set of works
cited\footnote{Specifically, the overlapping works cited are \citet{reith2002examination,technical2001electronic,carrier2003getting,beebe2005hierarchical,ciardhuain2004extended,cohen2010fundamentals,leigland2004formalization,ACPO2012,casey2010handbook}}
matches the \ac{ISO} work remarkably closely, as would be expected since
the primary authors are the same. Unfortunately, \citet{valjarevicanalyses}
gives absolutely no methodology for how they arrived at this list
of resources. Their analysis method is also not discussed, so it is
unclear how or why they arrived at their categories and classification.

These omissions are particularly strange in that \citet[p.~3]{valjarevicanalyses}
quotes \citet{cohen2011state} as rightly concluding the next steps
in reaching consensus on and improving the field of digital forensics
are a review of the literature that can be used to accurately drive
consensus. This task is clearly what's been attempted, and as it has
become an \ac{ISO} standard it seems to have been accepted by a variety
of practitioners. However, the lack of explanation of how these documents
were selected as the correct set from which to drive consensus makes
it hard to trace the authoritativeness of this source. 

\FloatBarrier

\subsubsection{FIRST Documents\label{sub:FIRST-referenced}}

\begin{table}[tb]
\begin{tabular}{|r|>{\centering}p{0.3cm}|>{\centering}p{0.3cm}|>{\centering}p{0.3cm}|>{\centering}p{0.3cm}|>{\centering}p{0.3cm}|}
\cline{2-6} 
\multicolumn{1}{r|}{} & \multicolumn{5}{c|}{Criteria}\tabularnewline
\hline 
Document & 1 & 2 & 3 & 4 & 5\tabularnewline
\hline 
\hline 
SP 800-53 \citep{nist800-53r4} & \ding{51} & \crossout{.1em}{.3cm} & \crossout{.1em}{.3cm} & \ding{51} & \ding{51}\tabularnewline
\hline 
SP 800-83 (\S4) \citep{nist800-83r1} & \ding{51} & \ding{51} & \ding{51} & \ding{51} & \ding{51}\tabularnewline
\hline 
SP 800-84 \citep{nist800-84} & \crossout{.1em}{.3cm} & \crossout{.1em}{.3cm} & \ding{51} & \ding{51} & \ding{51}\tabularnewline
\hline 
SP 800-86 \citep{nist800-86} & \ding{51} & \ding{51} & \ding{51} & \ding{51} & \ding{51}\tabularnewline
\hline 
SP 800-92 \citep{nist800-92} & \ding{51} & \ding{51} & \crossout{.1em}{.3cm} & \ding{51} & \ding{51}\tabularnewline
\hline 
SP 800-94 \citep{nist800-94} & \ding{51} & \crossout{.1em}{.3cm} & \crossout{.1em}{.3cm} & \ding{51} & \ding{51}\tabularnewline
\hline 
SP 800-115 \citep{nist800-115} & \ding{51} & \ding{51} & \crossout{.1em}{.3cm} & \ding{51} & \ding{51}\tabularnewline
\hline 
SP 800-128 \citep{nist800-128} & \ding{51} & \crossout{.1em}{.3cm} & \crossout{.1em}{.3cm} & \ding{51} & \ding{51}\tabularnewline
\hline 
\end{tabular}

\caption[List of \acs*{NIST} publications referenced by \citet{cichonski2012_sp800-61}]{\label{tab:NIST-referenced}\acs*{NIST} publications referenced by \citet{cichonski2012_sp800-61}.
The criteria are (1) target audience is security professionals; (2)
topic in scope, per Section~\ref{sec:Scope-Topic}; (3) focus is
investigator practices; (4) document finalized and not obsoleted as
of Aug 1, 2017; (5) available in English.}
\end{table}

\begin{table}[t]
\begin{tabular}{|r|>{\centering}p{0.3cm}|>{\centering}p{0.3cm}|>{\centering}p{0.3cm}|>{\centering}p{0.3cm}|>{\centering}p{0.3cm}|}
\cline{2-6} 
\multicolumn{1}{r|}{} & \multicolumn{5}{c|}{Criteria}\tabularnewline
\hline 
Document & 1 & 2 & 3 & 4 & 5\tabularnewline
\hline 
\hline 
\citet{mitre2010cyberscience} & \crossout{.1em}{.3cm} & \ding{51} & \crossout{.1em}{.3cm} & \ding{51} & \ding{51}\tabularnewline
\hline 
\citet{mundie2012building} & \crossout{.1em}{.3cm} & \ding{51} & \ding{51} & \crossout{.1em}{.3cm} & \ding{51}\tabularnewline
\hline 
\citet{fenz2009formalizing} & \crossout{.1em}{.3cm} & \ding{51} & \crossout{.1em}{.3cm} & \ding{51} & \ding{51}\tabularnewline
\hline 
\citet{osorno2011coordinated} & \ding{51} & \ding{51} & \ding{51} & \ding{51} & \ding{51}\tabularnewline
\hline 
\citet{magklaras2001insider} & \ding{51} & \crossout{.1em}{.3cm} & \ding{51} & \ding{51} & \ding{51}\tabularnewline
\hline 
\citet{wang2009ovm} & \ding{51} & \crossout{.1em}{.3cm} & \crossout{.1em}{.3cm} & \ding{51} & \ding{51}\tabularnewline
\hline 
\citet{chiang2009ontology} & \ding{51} & \crossout{.1em}{.3cm} & \crossout{.1em}{.3cm} & \ding{51} & \ding{51}\tabularnewline
\hline 
\citet{ekelhart2007security} & \ding{51} & \crossout{.1em}{.3cm} & \ding{51} & \ding{51} & \ding{51}\tabularnewline
\hline 
\citet{kossakowski1999responding} & \ding{51} & \ding{51} & \ding{51} & \ding{51} & \ding{51}\tabularnewline
\hline 
\end{tabular}

\caption[Documents referenced by \acs*{CERT/CC} sources]{\label{tab:Mundie-referenced}Documents referenced by the CERT documents.
The criteria are (1) target audience is security professionals; (2)
topic in scope, per Section~\ref{sec:Scope-Topic}; (3) focus is
investigator practices; (4) document finalized and not obsoleted as
of Aug 1, 2017; (5) available in English.}
\end{table}

As Table~\ref{tab:FIRST-results} demonstrates, we pass five \ac{FIRST}-related
documents through the evaluation of results to harvest further citations.
Three of these documents do not have any citations ready to harvest.
RFC~2196 \citep{rfc2196} does not have in-line citations, and only
\S5.4 is relevant, so the relevant citations to follow cannot be
distinguished. Further, RFC~2196 is already 20 years old, and so
following any citations would provide little modern benefit. On the
other hand, \citet{enisa2011proactive} is a primary source -- it
is a survey of preventative practices at over 100 \acp{CSIRT}. \citet{enisa2011proactive}
notes the tools that the respondents use, but it makes no citations
to other incident response methodology documents. \citet{alberts2004defining}
is similarly a primary source, though on incident management from
\ac{CERT/CC}. The only reference we take from \citet{alberts2004defining}
is where it explicitly indicates further information on incident analysis
is contained in another CERT document, namely \citet{kossakowski1999responding}.
Therefore, we harvest citations primarily from \citet{cichonski2012_sp800-61}
and \citet{mundie2014incident}. 

\citet{cichonski2012_sp800-61} references three classes of resources.
First is a list of incident response organizations, secondly a list
of \ac{NIST} publications related to incident response, and finally a list
of applicable data formats. The list of organizations includes many
already discussed in Section~\ref{sub:FIRST-results}. Those jointly
listed by \ac{NIST} and \ac{FIRST} are \ac{CERT/CC}, \ac{ENISA}, and \ac{FIRST} itself. \ac{NIST} additionally lists
the \ac{APWG}; \ac{CCIPS}; \ac{GFIRST}; \ac{HTCIA}; InfraGuard;
the \ac{ISC}; the National Council of \acp{ISAC}; and \ac{USCERT}. These organizations are
certainly involved in various aspects of incident response. However,
organizations as such are out of the scope of our review.\footnote{As a convenient sample, I have presented at \ac{APWG} \citep{spring2015global,spring2013modeling} and attended InfraGuard
meetings, and I do not expect there would be significant benefit in expanding the scope to include them. 
Likewise, I have
interacted with several \acp{ISAC}, and reviewed their available materials
(\ac{REN-ISAC} and \ac{FS-ISAC} especially) and do not believe they have any documents of importance to our topic.}

Table~\ref{tab:NIST-referenced} lists and evaluates the relevance
of the \ac{NIST} publications referenced by \citet{cichonski2012_sp800-61}.
All of these publications contribute to relevant background knowledge.
For example, any incident response professional will need to know
what a intrusion detection and prevention system is and how they are
deployed (SP 800-94). But this topic is not about evidence collection,
analysis, and reporting; it is merely necessary background knowledge.
The two publications that are relevant are the guides to \emph{Malware
Incident Prevention and Handling for Desktops and Laptops} \citep[\S4 only]{nist800-83r1}
and \emph{Integrating Forensic Techniques into Incident Response}
\citep{nist800-86}. 

The data exchange formats listed by \citet{cichonski2012_sp800-61}
are quite similar to those \ac{NIST}, \ac{IODEF}, and \ac{MITRE} formats extracted
from the \ac{IETF} documents in Table~\ref{tab:IETF-related-cite-acronyms}.
The only difference is the addition of Asset Identification (AI),
Asset Results Format (ARF), CVSS (from \ac{FIRST}, see Section~\ref{sub:FIRST-results}),
and Cyber Observable eXpression (CybOX). As discussed in Section~\ref{sub:IETF-referenced-docs},
these formats are languages for reporting results, but they do not
directly discuss what to say. Our intention is to discuss what to
say in results, while being language agnostic, which puts these various
formats and languages just out of scope. 

\citet{mundie2014incident} cites 27 documents. They include several
technical format documents for ontologies in the W3C Ontology Web
Language (OWL), KL-ONE knowledge representation, knowledge graphs,
process specification language (PSL), or the display tools used (Graphviz),
which we will not discuss. There are also psychology and ontology
that are obviously out of scope for our review \citep{miller1956magical,baader2003description}.
Further, four references we have already considered, namely \ac{ISO}/IEC
27001, \ac{ISO}/IEC 27002, \citet{cichonski2012_sp800-61}, and \citet{beebe2005hierarchical}.
These exceptions leave the eight documents evaluated in Table~\ref{tab:Mundie-referenced}.
The only documents that pass the evaluation are \citet{osorno2011coordinated}
and \citet{kossakowski1999responding}.

\citet{mitre2010cyberscience} on whether cybersecurity is a science
is not within our current, relatively narrow scope. However, since
incident response is an important subset of cybersecurity, whether
security investigations are a kind of subcategory of scientific investigation
clearly impacts our handling of what incident response is and
how to link it to knowledge generation and evidence evaluation more
generally. \citet{spring2017why} address the relationship between
cybersecurity and science and argue that cybersecurity as practiced
is a kind of science. 

\citet{fenz2009formalizing} provides a difficult decision. It is
one of the few attempts at formalization. However,
its target is security knowledge, not security practice.
This topic is closely allied to our hope to formalize incident response,
as security knowledge would be instrumental to that project. So while
not in scope for this review, this document may be useful for future
related work.

\subsubsection{IC Documents\label{sub:IC-referenced}}

\citet{heuer1999psychology} presents some challenges to adequate
reference harvesting. The book contains no collected list of references,
written in a traditional humanities style in which references are
in footnotes intermixed with commentary, but this is not the central
problem. As essentially a military intelligence and psychology book,
its sources are quite wide-ranging. References range from World War
II Nazi-propaganda analysis to behavioral economics. It is only through
Heuer's \ac{CIA} experience that these disparate sources are converted
into a useful guide on how to reason in adversarial situations. The
other challenge is that \citet{heuer1999psychology} makes only passing
reference to computers as tabulating machines. The closest he seems
to get to computer science is via \citet{simon1996sciences}, as he
discusses decision-making and satisficing. For these three reasons
we consider \citet{heuer1999psychology} as essentially a primary
source and do not trace citations from it. One should not be surprised
it has many features of a primary source, as surely its main value
is summarizing \ac{CIA} analytic experience not otherwise publicly available.

\citet{caltagirone2013diamond} and \citet{hutchins2011intelligence}
are more straightforward. There are 79 citations between the two,
with no overlap, though \citet{caltagirone2013diamond} cites both
\citet{hutchins2011intelligence} and \citet{heuer1999psychology}.
The references in \citet{caltagirone2013diamond} are noticeably more
strategy-focused over the tactically-focused \citet{hutchins2011intelligence},
as one would expect from their different topics. \citet{hutchins2011intelligence}
cites several vulnerability bulletins and company advisories as cases;
it is more of a primary source, documenting the analysis methods used
by Lockheed Martin incident response staff. We do not consider such
advisories, software tool documentation, and news items, as they are
not within our review topic. There are also several references already
covered elsewhere: \citet{cichonski2012_sp800-61}, STIX, CVE, and
SANS. There is even yet a new reporting and data exchange format:
Vocabulary for Event Recording and Incident Sharing (VERIS). We also
exclude references that are merely the official definitions of terminology.
These exceptions reduce the total referenced works to 50. 

The kill chain \citep{hutchins2011intelligence} and the diamond model \citep{caltagirone2013diamond} are both attack ontologies. They model the possible routes an adversary may take when executing an attack.  
One big class of cited work is other attack ontologies. We consider the kill chain and diamond model as \emph{de facto} standard ontologies, but we believe they reached that level of agreement within the IC because they also come with investigative norms for interpreting and filling in the ontologies.  
There are many other attack ontology works that do not come with such guidance, and so we will not consider them directly in scope for our review.  We remove 16 references from consideration because they are attack-ontologies without guidance.

There are also several references that are clearly for background or motivation, such as \citet{hawkins2004problem}, \citet{liu1998feature}, Symantec's analysis of the Duqu malware, and assessments of Chinese attack capabilities. We also remove how-to descriptions for conducting particular methods of technical analysis, namely on passive DNS analysis \citep{antonakakis2011Kopis}, crime-pattern analysis \citep{palmiotto1988crime}, and honeypots via the Honeynet Project. This leaves us with 22 documents in Table~\ref{tab:IC-referenced}. Of these, four pass our relevance requirements: \citet{cheswick1992evening},\footnote{Technically the citation is to this article's republication in a popular textbook, \citet[ch.~16]{cheswick2003firewalls}. However, as the rest of the textbook is not directly referenced, we reference just the original publication.} \citet{stoll1988stalking}, \citet{mitropoulos2006incident}, and \citet[ch.~5 only]{jp2-01-3}.

\begin{table}[t]
\begin{tabular}{|r|>{\centering}p{0.3cm}|>{\centering}p{0.3cm}|>{\centering}p{0.3cm}|>{\centering}p{0.3cm}|>{\centering}p{0.3cm}|}
\cline{2-6} 
\multicolumn{1}{r|}{} & \multicolumn{5}{c|}{Criteria}\tabularnewline
\hline 
Document & 1 & 2 & 3 & 4 & 5\tabularnewline
\hline 
\hline 
\citet{stamos2010aurora} & \ding{51} & \crossout{.1em}{.3cm} & \crossout{.1em}{.3cm} & \ding{51} & \ding{51}\tabularnewline
\hline 
\citet{amann2012lone} & \ding{51} & \crossout{.1em}{.3cm} & \crossout{.1em}{.3cm} & \ding{51} & \ding{51}\tabularnewline
\hline 
\citet{cheswick1992evening} & \ding{51} & \ding{51} & \ding{51} & \ding{51} & \ding{51}\tabularnewline
\hline 
\citet{stoll1988stalking} & \ding{51} & \ding{51} & \ding{51} & \ding{51} & \ding{51}\tabularnewline
\hline 
\citet{duran2009building} & \ding{51} & \crossout{.1em}{.3cm} & \crossout{.1em}{.3cm} & \ding{51} & \ding{51}\tabularnewline
\hline 
\citet{cohen1995protection} & \ding{51} & \crossout{.1em}{.3cm} & \crossout{.1em}{.3cm} & \ding{51} & \ding{51}\tabularnewline
\hline 
\citet{lewis2008holistic} & \crossout{.1em}{.3cm} & \crossout{.1em}{.3cm} & \crossout{.1em}{.3cm} & \ding{51} & \ding{51}\tabularnewline
\hline 
\citet{tirpak2000find} & \crossout{.1em}{.3cm} & \ding{51} & \ding{51} & \ding{51} & \ding{51}\tabularnewline
\hline 
\citet{hayes2008defending} & \ding{51} & \crossout{.1em}{.3cm} & \crossout{.1em}{.3cm} & \ding{51} & \ding{51}\tabularnewline
\hline 
\citet{willison2009overcoming} & \ding{51} & \ding{51} & \crossout{.1em}{.3cm} & \ding{51} & \ding{51}\tabularnewline
\hline 
\citet{mitropoulos2006incident} & \ding{51} & \ding{51} & \ding{51} & \ding{51} & \ding{51}\tabularnewline
\hline 
\citet{caltagirone2005adam} & \ding{51} & \crossout{.1em}{.3cm} & \ding{51} & \ding{51} & \ding{51}\tabularnewline
\hline 
\citet{caltagirone2005evolving} & \ding{51} & \crossout{.1em}{.3cm} & \ding{51} & \ding{51} & \ding{51}\tabularnewline
\hline 
\citet{bellovin1992there} & \ding{51} & \ding{51} & \crossout{.1em}{.3cm} & \ding{51} & \ding{51}\tabularnewline
\hline 
\citet{mcclure2005hacking} & \ding{51} & \crossout{.1em}{.3cm} & \ding{51} & \ding{51} & \ding{51}\tabularnewline
\hline 
\citet{brenner2002organized} & \crossout{.1em}{.3cm} & \crossout{.1em}{.3cm} & \crossout{.1em}{.3cm} & \ding{51} & \ding{51}\tabularnewline
\hline 
\citet{van1985electromagnetic} & \ding{51} & \crossout{.1em}{.3cm} & \crossout{.1em}{.3cm} & \ding{51} & \ding{51}\tabularnewline
\hline 
\citet{john2008detection} & \ding{51} & \crossout{.1em}{.3cm} & \ding{51} & \ding{51} & \ding{51}\tabularnewline
\hline 
\citet{jp3-13} & \ding{51} & \crossout{.1em}{.3cm} & \crossout{.1em}{.3cm} & {*} & \ding{51}\tabularnewline
\hline 
\citet{jp3-60} & \ding{51} & \crossout{.1em}{.3cm} & \ding{51} & {*} & \ding{51}\tabularnewline
\hline 
\citet[ch.~5 only]{jp2-01-3} & \ding{51} & \ding{51} & \ding{51} & {*} & \ding{51}\tabularnewline
\hline 
\end{tabular}

\caption[Resources referenced by \acl*{IC} documents]{\label{tab:IC-referenced}Documents referenced by \citet{caltagirone2013diamond}
and \citet{hutchins2011intelligence}. The criteria are (1) target
audience is security professionals; (2) topic in scope, per Section~\ref{sec:Scope-Topic};
(3) focus is investigator practices; (4) document finalized and not
obsoleted as of Aug 1, 2017; (5) available in English. The ({*}) in criterion four indicates we are referencing an updated version of the document. }
\end{table}

\FloatBarrier
\section{Discussion\label{sec:Discussion}}

Table~\ref{tab:Categorization-results} classifies advice on these
topics in several ways: to which phases the document applies, the
directness with which the document applies to each phase, the applicability
of the advice, investigative goals supported, broadness of scope,
generalizability of advice, and formalism. We shall make some commentary
on all the documents in the Table, in order of appearance, with one
exception. Before we lose ourself amongst the trees of this discussion,
we will make our general claim of the primary gap in the literature
clear up front. Our evocative evidence for this claim will be the
three \ac{NIST} publications, as they make the point most clearly. After
discussing these, we go on to discuss the \ac{ISO}, \ac{IETF}, etc. documents
in the order they appear in Table~\ref{tab:Categorization-results}. 

\begin{table*}[!th]
\begin{adjustwidth}{-1cm}{}

\begin{tabular}{|r|ccc|ccc|c|c|c|}
\cline{2-7} 
\multicolumn{1}{r|}{} & \multicolumn{3}{>{\raggedright}m{2cm}|}{ Directness per Phase} & \multicolumn{3}{>{\raggedright}m{2cm}|}{Scope\\

per Goal} & \multicolumn{1}{c}{} & \multicolumn{1}{c}{} & \multicolumn{1}{c}{}\tabularnewline
\cline{2-10} 
\multicolumn{1}{r|}{Document} & \multicolumn{1}{c|}{Col} & \multicolumn{1}{c|}{Anz} & Rep & \multicolumn{1}{c|}{Fix} & \multicolumn{1}{c|}{Int} & LE & General & Type & Formal\tabularnewline
\hline 
\hline 
27035-1:2016 \citep{iso27035-1-2016} & C & C & C & B & \xoutnovcm{.1em}{.6cm} & \crossout{.1em}{.6cm} & Un & Ont & Qual\tabularnewline
\hline 
27037 \citet{iso27037-2012} & D & \xoutnovcm{.1em}{.6cm} & \crossout{.1em}{.6cm} & \xoutnovcm{.1em}{.6cm} & \xoutnovcm{.1em}{.6cm} & M & Un & Ont & Qual\tabularnewline
\hline 
27041 \citep{iso27041-2015} & C & C & C & \xoutnovcm{.1em}{.6cm} & \xoutnovcm{.1em}{.6cm} & \crossout{.1em}{.6cm} & Likely & Instr & $\emptyset$\tabularnewline
\hline 
27042 \citet{iso27042-2015} & \xoutnovcm{.1em}{.6cm} & D & D & \xoutnovcm{.1em}{.6cm} & \xoutnovcm{.1em}{.6cm} & M & Un & Instr & $\emptyset$\tabularnewline
\hline 
27043\citep{iso27043-2015} & C & C & C & B & \xoutnovcm{.1em}{.6cm} & M & Un & Ont & Qual\tabularnewline
\hline 
RFC 2196, \S5.4 only \citep{rfc2196} & C & \xoutnovcm{.1em}{.6cm} & D & M & \xoutnovcm{.1em}{.6cm} & B & Likely & Instr & $\emptyset$\tabularnewline
\hline 
RFC 6545 \citep{rfc6545} & C & C & D & N & B & \crossout{.1em}{.6cm} & Un & Ont & Formal\tabularnewline
\hline 
RFC 7203 \citep{rfc7203} & C & \xoutnovcm{.1em}{.6cm} & C & N & N & N & Un & Ont & Formal\tabularnewline
\hline 
RFC 7970 \citep{rfc7970} & C & \xoutnovcm{.1em}{.6cm} & D & M & M & M & Un & Ont & Formal\tabularnewline
\hline 
RFC 8134 \citep{rfc8134} & C & \xoutnovcm{.1em}{.6cm} & \crossout{.1em}{.6cm} & B & B & \crossout{.1em}{.6cm} & Un & Study & $\emptyset$\tabularnewline
\hline 
\ac{NIST} 800-61 \citep{cichonski2012_sp800-61} & C & D & C & M & \xoutnovcm{.1em}{.6cm} & M & Un & Adv & Qual\tabularnewline
\hline 
\ac{NIST} 800-83 \S4 {\footnotesize{}\citep{nist800-83r1}} & D & \xoutnovcm{.1em}{.6cm} & C & B & \xoutnovcm{.1em}{.6cm} & \crossout{.1em}{.6cm} & Un & Ont & Qual\tabularnewline
\hline 
\ac{NIST} 800-86 \citep{nist800-86} & C & \xoutnovcm{.1em}{.6cm} & C & \xoutnovcm{.1em}{.6cm} & \xoutnovcm{.1em}{.6cm} & \crossout{.1em}{.6cm} & High & Study & Qual\tabularnewline
\hline 
\citet{enisa2011proactive} (\ac{ENISA}) & D & \xoutnovcm{.1em}{.6cm} & \crossout{.1em}{.6cm} & \xoutnovcm{.1em}{.6cm} & N & \crossout{.1em}{.6cm} & Un & Study & Qual\tabularnewline
\hline 
\citet{alberts2004defining} (\ac{CERT/CC}) & \xoutnovcm{.1em}{.6cm} & \xoutnovcm{.1em}{.6cm} & C & B & \xoutnovcm{.1em}{.6cm} & \crossout{.1em}{.6cm} & Un & Ont & Qual\tabularnewline
\hline 
\citet{kossakowski1999responding} (\ac{CERT/CC}) & C & D & C & M & \xoutnovcm{.1em}{.6cm} & \crossout{.1em}{.6cm} & Likely & Adv & Qual\tabularnewline
\hline 
\citet{mundie2014incident} (\ac{CERT/CC}) & C & C & C & B & \xoutnovcm{.1em}{.6cm} & \crossout{.1em}{.6cm} & High & Ont & Perf\tabularnewline
\hline 
\citet{osorno2011coordinated} (JHU \& US-CERT) & \xoutnovcm{.1em}{.6cm} & C & C & B & B & \crossout{.1em}{.6cm} & High & Ont & Qual\tabularnewline
\hline 
\citet{hutchins2011intelligence} (IC) & C & C & \crossout{.1em}{.6cm} & \xoutnovcm{.1em}{.6cm} & M & \crossout{.1em}{.6cm} & Un & Adv & Qual\tabularnewline
\hline 
\citet{caltagirone2013diamond} (IC) & C & D & \crossout{.1em}{.6cm} & \xoutnovcm{.1em}{.6cm} & M & \crossout{.1em}{.6cm} & High & Adv & Perf\tabularnewline
\hline 
\citet{heuer1999psychology} (\ac{CIA}) & \xoutnovcm{.1em}{.6cm} & D & \crossout{.1em}{.6cm} & B & B & B & Wide & Instr & Qual\tabularnewline
\hline 
\citet[ch.~5 only]{jp2-01-3} & \xoutnovcm{.1em}{.6cm} & D & \crossout{.1em}{.6cm} & \xoutnovcm{.1em}{.6cm} & M & \crossout{.1em}{.6cm} & High & Instr & Qual\tabularnewline
\hline 
\citet[ch.~2]{casey2010handbook} & C & D & D & \xoutnovcm{.1em}{.6cm} & \xoutnovcm{.1em}{.6cm} & B & Wide & Ont & Qual\tabularnewline
\hline 
\citet{mitropoulos2006incident} & C & C & C & M & N & \crossout{.1em}{.6cm} & Un & Study & Qual\tabularnewline
\hline 
\citet{carrier2004event} & D & C & D & \xoutnovcm{.1em}{.6cm} & \xoutnovcm{.1em}{.6cm} & M & Likely & Ont & Qual\tabularnewline
\hline 
\citet{ciardhuain2004extended} & C & C & C & \xoutnovcm{.1em}{.6cm} & \xoutnovcm{.1em}{.6cm} & B & Un & Ont & Perf\tabularnewline
\hline 
\citet{leigland2004formalization} & D & \xoutnovcm{.1em}{.6cm} & \crossout{.1em}{.6cm} & N & N & N & Likely & Instr & Formal\tabularnewline
\hline 
\citet{stoll1988stalking} & C & D & C & N & M & N & Likely & Study & $\emptyset$\tabularnewline
\hline 
\citet{cheswick1992evening} & C & D & \crossout{.1em}{.6cm} & \xoutnovcm{.1em}{.6cm} & B & \crossout{.1em}{.6cm} & Likely & Study & $\emptyset$\tabularnewline
\hline 
\end{tabular}

 \end{adjustwidth}

\caption[Results: document categorization]{\label{tab:Categorization-results} Categorization of relevant documents. 
The phases are collection, analysis, and reporting. 
Cells have a cross if a phase is not addressed.
Advice directness values are direct (D) or constraints-based (C). 
Values for goals are to fix an infected system (fix), gathering intelligence (int), and law enforcement (LE) action.
Values for a document's intended scope are narrow (N), medium (M), or broad (B). 
Values for generalizability of an approach are unlikely (Un), likely (Likely), highly likely
(High), or already widely generalizable (Wide). 
Document types are case studies (Study), ontologies (Ont), advice on actions (Adv), and explicit instructions (Instr). 
Values for formalization are not present ($\emptyset$), qualitative (Qual), formal, or perfunctory (Perf).}
\end{table*}

Despite recommending everyone use a methodical approach, \ac{NIST} fails
to provide one. This failure is symptomatic of the state of available
practicable policy advice and practitioner training material. This
is the central gap identified by the literature review: There may
be adequate concrete tools and training available, but there is no
general training or policy advice for strategic selection of tactics,
that is,\textbf{ which analysis heuristic or technical tool to employ
in a particular situation and why}. An attendant gap is a failure
to advise on \textbf{when the investigator is justified in generalizing};
that is, making a stronger, broader claim from singular pieces of
evidence. Because there is no advice on which strategy to employ,
or when broadening claims are justified, there is similarly a gap
in \textbf{what information to report in order to convince a reader
that the investigator should be believed}. 

Upon glancing through \ac{NIST} SP 800-61 \citep{cichonski2012_sp800-61}
it is obvious why all four of our venues reference it or use it as
their standard directly. It is comprehensive and thorough without
being overbearing. However, its focus is incident management, not
investigation. The analysis phase receives about three pages of discussion
(p 28-30), reporting one page (p 31),\footnote{SP 800-61 acknowledges its discussion of reporting is too brief, and
refers the reader to RFC 5070. This document has since been obsoleted
by RFC 7970 \citep{rfc7970}.} and evidence collection half a page (p 36), and general decision
making and prioritization two pages (p 32-33). \citet[p~32]{cichonski2012_sp800-61}
addresses the problem of scarce resources directly: ``prioritizing
the handling of the incident is perhaps the most critical decision
point in the incident handling process.'' The following discussion,
while short, is two more pages about decision-making during incident
analysis than almost any other document found. Regardless, it is not
sufficient to develop a robust account of the nuances and difficulties
an investigator regularly deals with when evaluating evidence, generalizing
from particulars, and deciding how best to report their findings. 

\ac{NIST} SP 800-83 \citep{nist800-83r1} \S4.2 is titled ``detection
and analysis,'' yet we have provocatively labeled the document has
having no bearing on the analysis phase. The section's advice on analysis
is, however, entirely tool-focused pragmatics. Analysis should take
place on an isolated or virtualized operating system to prevent spread
of infection, and so on. The document mentions some fields that the
investigator may want to collect, such as file names, service ports,
and ``how to remove the malware.'' There is no advice on how to
obtain this information, why, or what it might be useful for. Therefore,
these are best understood as reporting constraints, not analysis advice.
This result is disappointing, considering \S4 gives its opening motivation
as ``this section of the guide builds on the concepts of SP 800-61
by providing additional details about responding to malware incidents.'' 

\ac{NIST} SP 800-86 \citet{nist800-86} suffers similarly to SP 800-83;
it consists of a stream of data formats and types and assumes that
the investigator will know what to do know that the possible data
types have been listed. These make up underlying technical skills
necessary for an investigation, and so are not completely irrelevant
to incident response. However, they do not help us understand
how investigators make decisions. The extent of the advice on analysis
again amounts to essentially a reporting and collection constraints,
respectively: ``the analysis should include identifying people, places,
items, and events, and determining how these elements are related...
often, this effort will include correlating data among multiple sources''
\citep[p~3-6]{nist800-86}. 

These \ac{NIST} documents provide a useful effort to summarize our gaps
and intended way forward. \citet[p~3-8]{nist800-86} recommends all
organizations have an incident response capability and that analysts
use ``a methodical approach to a digital forensic investigation.''
The document clearly states the importance of digital forensic investigation
and advises on terminology, analysis techniques, and pitfalls to avoid.
Despite \ac{NIST}'s policy recommendation to do so, based on this survey
\ac{NIST} does not actually provide a methodical approach to analyzing
data during an investigation. SP 800-61 comes closest, but the discussion
of analysis method there still amounts to an unordered collection
of tips, tricks, and pitfalls to avoid. While these are all accurate
and sound advice, they do not comprise a method. The \ac{NIST} documents
evaluated and surveyed here are a generally positive attempt at providing
practical advice to a wide audience on a complex topic. Their focus
is practical guides on aspects of tools used in digital forensics.
But the assumption is that once an investigator is taught how to use
a tool, they will know when and why to use it. This gap is a recurring
assumption, and precisely our intended focus for improvement.

\FloatBarrier

\subsection{Comments and clarifications on Table~\ref{tab:Categorization-results}}
\begin{description}
\item [{The ISO 27000-series}] is dedicated to information security.
We have identified five standards on that are within our scope of
the particular parts of incident response. The relationship between
these standards is documented by Figure~1, reproduced in each of
the \ac{ISO} standards. This figure states clearly that all the listed
standards are applicable to ``investigation process classes and activities''
\citep[p.~ix]{iso27043-2015}. Process classes are readiness, initialization,
acquisitive, and investigative; activities overlap these classes,
and are plan, prepare, respond, identify-collect-acquire-preserve,
understand, report, and close. This taxonomy is essentially consistent
with the taxonomies used by the \ac{IETF}, \ac{NIST}, and \ac{FIRST} \citep{mundie2014incident}.
\\
However, where a \ac{NIST} standard such as SP 800-61 is a single 70-page
document, the \ac{ISO} incident response standards are each 10-15 pages
of unique content with 10-15 pages that are repeated in each document.
Thus, the five \ac{ISO} documents combined are comparable in scope and
detail to SP 800-61. However, unlike a \ac{NIST} publication, the \ac{ISO} documents
do not present a clear investigative goal among the options we have
distinguished, even within documents, let alone among them. 
\item [{27043,}] for example, seems to unknowingly alternate between incident
response for fixing systems and analysis for providing evidence to
a legal proceeding. Within 27043 \citet{iso27043-2015}, \S8 reads
like advice from \ac{CERT/CC} \citep{alberts2004defining}, and \S9 reads
like advice from \citet[ch.~2]{casey2010handbook}. The shift is abrupt
and without explanation. The shift includes a shift in terminology
and jargon for referring to essentially the same mental process by
the investigator. This oddity does not build confidence that the \ac{ISO}
standards actually present a unified methodology for incident response
as a series of disconnected vignettes. 
\item [{27041}] does little to dispel this sense of disconnectedness. This
\ac{ISO} document is disconnected from the other incident management documents
in that it focuses on the client-contractor relationship. The sense
in which is a process is validated is that ``the work instruction
\foreignlanguage{british}{fulfils} the requirements agreed with the
client'' \citep[p~9]{iso27041-2015}. 27041 \citet[p~12-13]{iso27041-2015}
states an investigation composed of validated examinations ``can
be considered to be validated'' while defining a validated examination
as one mode up of validated processes. Assuming composability of valid
processes is a dangerous claim. Concurrent program verification has
shown such claims cannot be assumed and are challenging to prove \citep{milner1989communication,ohearn2007resources};
albeit the technical sense of ``valid'' is slightly different, doubt
in the \ac{ISO} assumption seems warranted.
\end{description}
For these reasons, the \ac{ISO} standards would struggle to function well
as a unified whole. There does not seem to be an overarching editorial
guidance to assure consistency or navigate conflicts. At best, if
the reader already knows how to navigate the different, conflicting
contexts, the \ac{ISO} documents are useful expressions of each area of
concern. The level of detail is appropriate for ensuring management
ability to oversee a process, rather than to do the process itself.
Even the most specific documents (27037 and 27042), to which the other,
more general documents refer for details, are thin on anything that
might help with actual decision-making. \ac{ISO}/IEC 27042 provides a basic
distinction between static and dynamic analysis of malware (it uses
``live'' for dynamic), but all that is really provided are a few
descriptions of what distinguishes static and dynamic analysis. These
descriptions do not provide information on how to actually do either
kind of analysis, or even common pitfalls or errors to avoid. 
\begin{description}
\item [{RFC 2196}] is quite old, and its advice shows its age. The steps
are in general sound; however, they are from a time when it was reasonable
to ask for ``all system events (audit records)'' to be recorded
and evaluated by the investigator \citep[p~54]{rfc2196}. The text
assumes that incident investigators will know what to do with these
events once logged. This advice is not bad, such as it is; however
it is best understood as historical rather than actionable advice. 
\item [{RFC 6545}] and RFC 6546 jointly detail Real-time Inter-network
Defense (RID). RFC~6545 describes conceptual and formal details,
whereas RFC~6546 provides technical communication and encryption
details. RFC~6545 is an extension of \ac{IODEF} \citep{rfc7970}, specifying
methods and norms of communication using \ac{IODEF} between organizations.
As such, the document focuses on what to report, and how to use reports
for mitigation. Policy of use and sensitivity of information is explicitly
integrated into the format. How analysis produces adequate data is
out of scope. However, by providing such explicit standards on what
should be reported and how those reports can expect to be used, RID
does put constraints on analysis and evidence collection -- those
phases need to produce reporting with the specified types of fields. 
\item [{RFC 7203}] extends \ac{IODEF} \citep{rfc7970} ``to embed and convey
various types of structured information'' \citet[p~2]{rfc7203}.
Specifically, the various metrics and formats such as CVSS and CVE
captured in Table~\ref{tab:IETF-related-cite-acronyms}. This extension
serves to integrate two types of reporting format and constraint.
This is useful, but is mostly programmatic. Therefore, it is not directly
about reporting in the same way RFC~7970 is. Although technically
detailed, from a decision-making point of view RFC~7203 just suggests
that these metrics are useful ways to describe an incident and report
on it, and that investigators should do so. RFC~5901 makes similar
suggestions specifically for reporting phishing \citep{rfc5901}. 
\item [{RFC 7970}] is the heart of the \ac{IETF} incident analysis standardization
effort. It obsoletes RFC 5070, which is cited or used by most publication
venues as the incident reporting format. The focus is on exchanging
indicators of incidents for collective defense. Although \ac{IODEF} is,
strictly speaking, just an XML schema for document incidents, the
available options and the ontology provided to constrain the other
phases up to reporting. For some fields, this provides only minimal
collection requirements. However, consider the system impact type
attribute, which is a required field. There are 24 options specified,
ranging from ``takeover-account'' to ``integrity-hardware'' \citep[p~46]{rfc7970}.
Individuating among these various impacts would require a relatively
sophisticated incident response and analysis capability; it is
not so easy as logging an IP address and passing it along. Just within
the assessment class, one of two dozen overarching classes, there
are five types of impact to distinguish between with similar detail:
system, business, time, money, and confidence. Such detail provides
the most rigorous reporting requirements and guidance available. 
\item [{RFC 8134}] is informational, and not a standard. It provides a
list of information exchanges, collaborations, and implementations
that make use of \ac{IODEF}. Because information exchanges are a source
of evidence collection, the details about what information is available
from what groups provides evidence collection suggestions and introductions.
Although this advice at a rather abstract level, it is useful because
it provides a discussion of network defense information sharing arrangements
that is not commonly quite so public. 
\item [{\citet{enisa2011proactive}}] is a study commissioned by \ac{ENISA}
and executed by the Polish CERT. The focus is on data sources -- how
do \acp{CSIRT} monitor their constituents. The method employed is a survey
of over 100 \acp{CSIRT}. While this data is at best instructive of where
to get data, it is an important resource for how respondents evaluate
the quality of data sources. Such evaluation is directly relevant
to evidence collection decisions. It is unlikely this study is instructive
outside this relatively narrow context. However, it is directly relevant
context for this work. 
\item [{\citet{alberts2004defining}}] is primarily about contextualizing
incident management within a wider organizational context. In fact,
\citet[pp~24-26]{alberts2004defining} is one of the best assessments
of the relationship of investigation to preparation and protection
we have found in this review. However, our focus is not on situation
of the investigative process within an organization. \citet[p~128ff.]{alberts2004defining}
is an ambitious effort to organize a flow chart for incident response.
Because their scope includes technical, management, and legal responses,
the level of detail devoted to analysis amounts to ``Designated personnel
analyze each event and plan, coordinate, and execute the appropriate
technical response'' \citep[p~136]{alberts2004defining}. 
\item [{\citet[p 17ff.]{kossakowski1999responding}}] provides classes
of advice, like collect logs and isolate infected machines from the
network. These perhaps come the closest to advice about how to collect
evidence from computer incidents. However, it is silent on which logs
to collect, or what to look for when examining network traces. While
this advice is highly likely to be able to generalize to all cases
of incident response, the level of detail is not operationalizable
as a decision-making instructions. 
\item [{\citet{mundie2014incident}}] is, in effect, a literature review
of incident management. As such, it mostly constrains the inputs and
outputs one would expect from incident response. The formalism
provided is in a specification of an OWL ontology language of incident
management. This language is a useful step in reconciling various
incident management processes. However, it is simply a few levels
of abstraction above our current task.
\item [{\citet{osorno2011coordinated}}] has done something similar to
our project here, in that they inventory various incident management
processes, with two main differences. They focus on moving up a level
to inter-organizational coordination during complicated incidents,
rather than zooming in on individual analyst decision-making. \citet{osorno2011coordinated}
also focuses on the US context. This different purpose leads to substantial
differences in emphasis as to what is reviewed; for example, where
we have generally set aside data exchange formats (see Section~\ref{sub:IETF-referenced-docs}),
\citet{osorno2011coordinated} spend considerable effort mapping these
formats into each other. For this reason, the extent of their recommendation
on incident response amounts to do an OODA-style loop, a military
term standing for observe, orient, decide, and act \citep[p~7]{osorno2011coordinated}. 
\item [{\citet{hutchins2011intelligence}}] discusses courses of action
for network defense. However, these are not direct advice on incident
investigation steps. The level of advice is on the order of ``to
disrupt installation events, use anti-virus.'' This advice is sound,
but it is not particularly concrete. 
\item [{\citet{caltagirone2013diamond}}] provides some light formalization
of their qualitative categories into both graph theory and subject
probabilities and Bayesian statistics. We call this perfunctory formalization
because the extent of their documentation is to list the formal structures
that are equally well-described by their prose. The document does
not make any use of the formalism, and it is not central to their
arguments. The structures appear to be constructed adequately; it
is simply that any application of them is left as an exercise for
the reader in much the same way as if they were not there. 
\item [{\citet{heuer1999psychology}}] is an instructive case for some
advice being too broad for our purposes. The book is comprised of
explicit decision-making instructions for analysis of intelligence.
However, the level of abstraction is so broad that it can be argued
to be applicable to almost any adversarial decision-making environment.
So while it is valuable, and it provides instructions for avoiding
cognitive biases, it does not provide instructions at the level of
detail that are directly useful for an investigation. 
\item [{\citet[ch. 5]{jp2-01-3}}] is about how to think like your adversary.
It is an extended treatment of developing and evaluating adversary
action plans across multiple dimensions under constrained resources.
The basic cycle is to identify objectives, enumerate courses of action,
evaluate and rank the likelihood of following each action, and identify
necessary intelligence collection requirements to determine adversary
decisions. The document is about military intelligence operations
generally, not computer-security incidents. However, one narrow but
necessary aspect of any investigation is how to anticipate an adversary.
This document covers the thought process behind the topic of anticipation
in a way which should be easily applicable to computer security. 
\item [{\citet[ch. 2]{casey2010handbook}}] is built around the claim that
digital forensics is just another kind of scientific investigation.
The basic ontology of the scientific method is represented as simply
create and evaluate hypotheses dispassionately based on evidence.
This description is supported by several case studies as examples
working through the method. This pedagogical strategy does not quite
amount to what we call advice on decision making, but it is also more
than merely an ontology. The end of the chapter also includes advice
on how to report conclusions convincingly to a jury or prosecuting
attorney. The advice is simple, but direct and effective: be concise,
let the evidence speak for itself, do not jump to conclusions, summarize
key results up front. The target audience is law enforcement who will
be using information technology to support general legal cases, not
computer crimes. Despite this broad audience, the treatment of decision-making
as part of the scientific method allows for easy and broad re-purposing
for other scenarios more directly related to computer-security incidents. 
\item [{\citet{mitropoulos2006incident}}] is relevant to our project for
its series of small case studies on how to analyze different types
of network logs. First, they provide a flow chart of analysis of a
IDS logs, intended to serve as an example of analysis of a particular
kind of technical information. Although only represented at a relatively
abstract level, this is directly a representation of reasoning and
decision making during incident response. They also provide basic
instructions on how to gather intelligence data on adversaries using
various network protocols. While we have ranked this as unlikely to
generalize, it is also very specifically targeted to techniques that
are commonly useful during incident response. So it is not particularly
necessary that they are generalizable. These sorts of cases are difficult
to teach and capture in a more abstract form, and are inflexible,
and so this specificity does come with a cost. Another important difficulty
is that \citet{mitropoulos2006incident} do not advise in which scenarios
to employ these different types of forensic techniques. 
\item [{\citet{carrier2004event}}] describes a evidence collection and
hypothesis testing and reporting model of digital forensic investigation.
They use forensic in its formal legal sense, and so target specifically
the gathering of adequate legal evidence. Their treatment of analysis
is particularly brief, and roll it into part of the evidence collection
process as simply a determination of if the data element is relevant
to the defined target. As to target definition, it ``is the most
challenging of the search phase'' and is done ``from either experience
or existing evidence'' \citep[p~8]{carrier2004event}. Like the other
documents, this one sidesteps most of the actually hard work on which
we would like to focus as essentially out of their scope. 
\item [{\citet{ciardhuain2004extended}}] attempts a novel formalization
by incorporating information flow; although not cited, this is a term
likely taken from \citet{barwise1997information}. However, the application
to forensic investigation \citep[p~21]{ciardhuain2004extended} bears
little resemblance to formal information flow models. The discussion
places vague constraints on evidence reporting such as ``the investigators
must construct a hypothesis of what occurred'' and ``The hypothesis
must be presented to persons other than the investigators'' \citep[p~7]{ciardhuain2004extended}.
But in practice these are of little use in making decisions during
an investigation. 
\item [{\citet{leigland2004formalization}}] provides a formal specification
of evidence collection methods that can be adapted to specific operating
systems. The language associates collection goals with certain common
attack patterns. The goal is narrowly practical -- to speed collection
of evidence by technicians during an investigation while reducing
superfluous data collection to make analysis a bit easier. The language
maps general actions to specific operating-system commands. The technician
needs to specify file identifiers for specific attack campaigns; I
expect the main downfall of this method is that adversaries learned
to randomize certain identifiers within their attacks. Such randomization
makes keeping an adequate library of definitions for this language
as defined essentially impossible. 
\end{description}

\subsection{Note on case studies }

Case studies or collections of cases that have been analyzed by others
provide demonstrations of what sort of attacks are possible. We have
two of the earliest examples of this style of reporting with \citet{stoll1988stalking}
and \citet{cheswick1992evening}. It is important to note these are
examples. A survey of incident case studies may be a useful additional
project, though it is out of our scope. The practitioners who wrote
these standards documents would be aware that many security vendors
publish accounts of adversaries they have tracked in the course of
their work. These are of varying quality, scope and importance. More
recent impactful studies include, for example, Mandiant tracking an
alleged unit of the Chinese military~\citep{mandiant2013apt1} and
Google's self-report of comprise attributed to China~\citep{google2010aurora}.
Some case reports are official government commissioned, such as the
Black Tulip report analyzing the compromise of a TLS certificate authority
critical to the operation of many Dutch government websites~\citep{foxit2012diginotar}. 

The scope need not be an individual case. Some studies focus on trends
rather than individual cases. Verizon's Data Breach Investigation
Report is probably the best-known example (see, e.g., \citet{verizon2015dbir,verizon2016dbir}).
United States federal civilian agencies must make annual breach reports
to Congress per FISMA requirements; such detailed reports have been
examined for trends \citep{pang2017security}.

One notably change in this style of report since \citet{stoll1988stalking}
and \citet{cheswick1992evening} is a trend away from discussing how
exactly the investigators found what they found. In an environment
where adversaries are likely to read any reports their targets publish,
this shift towards withholding information is understandable. Paradoxically,
this makes the old case studies more valuable, as they remain some
of the better expressions of the investigator's thought process. Of
course, the tools and networks the old case studies discuss are almost
entirely irrelevant, which can make them hard to apply to today's
systems. And the case studies do not do any of the work to make the
necessary generalizations. Through our survey, we find many of the
modern expressions of the form of incident response cycles are
quite consistent with the mental process \citet{stoll1988stalking}
and \citet{cheswick1992evening} describe.

\section{Conclusion\label{sec:IR-conclude}}

Our review of the incident analysis literature indicates a gap specifically around decision-making. There is adequate advice at a management level and a technical operational instructions level. But no adequate advice was found for decisions at a middle-level of granularity; specifically, gaps of note are:
\begin{itemize}
\item strategic selection of tactics, that is, which analysis heuristic or technical tool to employ in a particular situation and why
\item when the investigator is justified in generalizing; that is, making a stronger, broader claim from singular pieces of evidence
\item what information to report and how to communicate it in order to convince someone that the investigator should be believed
\end{itemize}

These problems are similar to those that scientists face in conducting their research. 
Incident responders have quite different operating environments that most scientists. 
However, as argued by \citet{spring2017why}, there is no obvious barrier to considering security research as a type of science. 
Scientific methods and norms need to be adapted, as security poses certain novel challenges.
However, as argued by \citet{spring2018generalization}, security analysts already take an approach to generalizing knowledge that is both similar to and can benefit from the wider literature on scientific explanation in philosophy of science.
Therefore, it is plausible that incident response would benefit from answering these gaps with methodology adapted from philosophy of science, though it is unlikely to be a panacea. 

Various other disciplines might contribute to filling these decision-making gaps. 
Reviews of such other fields are future work. 
Note that none such arose through our review of incident response standards. 
This gap indicates a reticence to take on scientific tools. 
The exception is the \acl{IC}, which genuinely integrates psychology and behavior economics \citep{heuer1999psychology} and to some extent Bayesian belief propagation \citep{caltagirone2013diamond}.
Other fields that might naturally be interrogated for links to incident analysis in future work include game theory, decision theory, information theory, systems engineering, internet measurement, and risk assessment.
As one example, game theory and network security already have a developed overlap, for example see \citet{alpcan2011network}, which might be adaptable to incident response. 

However, we are not aware of any work that has attempted to formalize decision-making in incident response specifically. 
We leave this as an area of future work.    
The work by \citet{horneman2017how}, adapting \citet{heuer1999psychology} to computer network analysis, is the closest approach so far. 
However, when our review calls for more structure to decision-making, it would mean further structure and specification of what she identifies as ``analytical acumen'' and its use.  

Finally, we would like to note that this lack of published standards for this granularity of decision-making does not mean that all incident responders are ignorant or hopeless. 
Incident analysis is trade-craft, essentially handed down by apprenticeship. 
Much more likely, various norms of reasoning through the three gaps we note have been developed and disseminated amongst small groups of analysts. 
Getting access to and surveying these analysts and their reasoning methods would also be a rich area for future work, though it is fraught with difficulties of gaining trust, access, time, and representative samples. 
The abundance of incident response teams, and the general social importance of responding to computer security incidents, has perhaps made this trade-craft approach unsustainable. 
The unsustainable nature of such trade-craft approaches heavily influences our conviction that it is time to consolidate decision-making in incident analysis and begin publicly filling in these gaps.

\section*{Acknowledgements}
Spring was supported by University College London's Overseas Research Scholarship and Graduate Research Scholarship.
orcid.com/0000-0001-9356-219X

\printbibliography

\pagebreak[1] %

\subsection*{List of Acronyms}
\label{sec:acronyms}
\begin{acronym}[jspring-thesis]
        \acro{ACM}{Association for Computing Machinery}
    	\acro{ACoD}{\acroextra{Art into Science: }A Conference for Defense}
    	\acro{AES}{Advanced Encryption Standard}
    	\acro{AirCERT}{Automated Incident Reporting}
    	\acro{APWG}{Anti-Phishing Working Group}
	\acro{ARMOR}{Assistant for Randomized Monitoring Over Routes}
    	\acro{ARPA}{Advanced Research Projects Agency\acroextra{, from 1972--1993 and since 1996 called \acs{DARPA}}}
	\acro{attack}[ATT\&CK]{Adversarial Tactics, Techniques, and Common Knowledge\acroextra{ (by \acs{MITRE})}}
    	\acro{BCP}{Best Current Practice\acroextra{, a series of documents published by \acs{IETF}}}
    	\acro{BGP}{Border Gateway Protocol}
    	\acro{BI}{logic of bunched implications}
    	\acro{BIS}{Department for Business, Innovation, and Skills\acroextra{ (United Kingdom)}}
	\acro{BLP}{Bell-Lapadula\acroextra{, a model of access control}}
    	\acro{CAE}{Center of Academic Excellence}
    	\acro{CAIDA}{Center for Applied Internet Data Analysis\acroextra{, based at University of California San Diego}}
    	\acro{CAPEC}{Common Attack Pattern Enumeration and Classification\acroextra{ (by \acs{MITRE})}}
    	\acro{CCIPS}{Computer Crime and Intellectual Property Section\acroextra{ of the \ac{DoJ}}}
    	\acro{CCE}{Common Configuration Enumeration\acroextra{ (by \acs{NIST})}}
    	\acro{CCSS}{Common Configuration Scoring System\acroextra{ (by \acs{NIST})}}
    	\acro{CEE}{Common Event Expression\acroextra{ (by \acs{MITRE})}}
    	\acro{CERT/CC}{CERT{\small{}{\textregistered}} Coordination Center\acroextra{ operated by Carnegie Mellon University}}
	\acro{CIA}{Central Intelligence Agency\acroextra{ (\ac{US})}}
    	\acro{CIS}{Center for Internet Security}
        \acro{CNA}{Computer Network Attack}
    	\acro{CND}{Computer Network Defense}
    	\acro{CNO}{Computer Network Operations}
    	\acro{CPE}{Common Platform Enumeration\acroextra{ (by \acs{NIST})}}
    	\acro{CSIR}{Computer Security Incident Response}
    	\acro{CSIRT}{Computer Security Incident Response Team}
    	\acro{CTL}{Concurrent Time Logic}
    	\acro{CVE}{Common Vulnerabilities and Exposures\acroextra{ (by \acs{MITRE})}}
    	\acro{CVRF}{Common Vulnerability Reporting Framework}
    	\acro{CVSS}{Common Vulnerability Scoring System\acroextra{, maintained by \acs{FIRST}}}
    	\acro{CWE}{Common Weakness Enumeration\acroextra{ (by \acs{MITRE})}}
    	\acro{CWSS}{Common Weakness Scoring System\acroextra{, maintained by \acs{MITRE}}}
    	\acro{CybOX}{Cyber Observable Expression\acroextra{, maintained by \acs{MITRE}}}
    	\acro{DARPA}{Defense Advanced Research Projects Agency}
    	\acro{DHS}{\ac{US} Department of Homeland Security}
    	\acro{DNS}{Domain Name System}
    	\acro{DoD}{\ac{US} Department of Defense}
    	\acro{DoJ}{\ac{US} Department of Justice}
    	\acro{ENISA}{\acs{EU} Agency for Network and Information Security}
    	\acro{EPSRC}{Engineering and Physical Sciences Research Council\acroextra{ (United Kingdom)}}
    	\acro{EU}{European Union}
    	\acro{FAA}{Federal Aviation Administration\acroextra{ (\ac{US})}}
	\acro{FBI}{\acroextra{\ac{US} }Federal Bureau of Investigation}
	\acro{FDA}{\acroextra{\ac{US} }Food and Drug Administration}
    	\acro{FIRST}{Forum of Incident Response and Security Teams}
    	\acro{FISMA}{Federal Information Security Management Act\acroextra{ (\ac{US})}}
    	\acro{FS-ISAC}{Financial Services \acf{ISAC}}
    	\acro{GCHQ}{Government Communications Headquarters\acroextra{ (United Kingdom)}}
    	\acro{GFIRST}{Government \acs{FIRST}}
    	\acro{HotSoS}{Symposium on the Science of Security}
    	\acro{HTTP}{Hypertext Transfer Protocol\acroextra{, a standard by \acs{W3C}}}
    	\acro{HTCIA}{High Technology Crime Investigation Association}
    	\acro{IC}{intelligence community}
    	\acro{ICT}{information and communications technology}
    	\acro{IEEE}{Institute of Electrical and Electronic Engineers}
    	\acro{IEP}{Information Exchange Policy}
    	\acro{IETF}{Internet Engineering Task Force}
    	\acro{IDS}{intrusion detection system}
    	\acro{IODEF}{Incident Object Description Exchange Format}
    	\acro{iodefplus}[IODEF+]{Incident Object Description Exchange Format Extensions\acroextra{ (RFC~5901)}}
    	\acro{IDMEF}{Intrusion Detection Message Exchange Format\acroextra{ (RFC~4765)}}
    	\acro{ISAC}{Information Sharing and Analysis Center}
    	\acro{ISC}{Internet Storm Center\acroextra{part of the privately-run \acs{SANS}}}
    	\acro{ISO}{International Organization for Standardization}
	\acro{ISP}{Internet Service Provider}
    	\acro{ITU}{International Telecommunications Union\acroextra{, an agency of the \acs{UN}}}
    	\acro{LAX}{Los Angeles International Airport}
    	\acro{LBNL}{Lawrence Berkeley National Laboratory}
    	\acro{MAEC}{Malware Attribute Enumeration and Characterization\acroextra{ (by \acs{MITRE})}}
    	\acro{MITRE}{the Mitre Corporation}
    	\acro{MMDEF}{Malware Metadata Exchange Format}
    	\acro{MoD}{Ministry of Defence\acroextra{ (United Kingdom)}}
    	\acro{NATO}{North Atlantic Treaty Organization}
	\acro{NCA}{National Crime Agency\acroextra{ (UK)}}
    	\acro{NCCIC}{\acroextra{\ac{US} }National Cybersecurity and Communications Integration Center}
    	\acro{NDA}{non-disclosure agreement}
    	\acro{NIDPS}{Network Intrusion Detection and Prevention System}
    	\acro{NIST}{National Institute of Standards and Technology\acroextra{, part of the \ac{US} Department of Commerce}}
	\acro{NSA}{National Security Agency\acroextra{ (\ac{US})}}
    	\acro{NSF}{National Science Foundation\acroextra{ (\ac{US})}}
    	\acro{OCIL}{Open Checklist Interactive Language\acroextra{ (by \acs{NIST})}}
    	\acro{OVAL}{Open Vulnerability and Assessment Language\acroextra{ (by \acs{MITRE})}}
    	\acro{OWASP}{Open Web Application Security Project}
    	\acro{OWL}{Ontology Web Language}
    	\acro{pDNS}{passive \acf{DNS}\acroextra{ traffic analysis}}
    	\acro{RAM}{Random Access Memory}
    	\acro{RCT}{Randomized Controlled Trial}
    	\acro{REN-ISAC}{Research and Education Networking \acf{ISAC}}
    	\acro{RID}{Real-time Inter-network Defense}
    	\acro{RISCS}{Research Institute in Science of Cyber Security\acroextra{ (United Kingdom)}}
    	\acro{RFC}{Request for Comments\acroextra{, standardization and informational documents published by the \ac{IETF}}}
    	\acro{SANS}[SANS Institute]{Sysadmin, Audit, Network, and Security Institute}
    	\acro{SCAP}{Security Content Automation Protocol\acroextra{ (by \acs{NIST})}}
    	\acro{SiLK}{System for Internet-level Knowledge\acroextra{, an open-source analysis tool set published by \ac{CERT/CC}}}
    	\acro{SoK}{Systematization of Knowledge\acroextra{ paper in \acs{IEEE} Oakland conference}}
	\acro{STIX}{Structured Threat Information Expression\acroextra{ (by \acs{MITRE})}}
    	\acro{STS}{Science and Technology Studies\acroextra{ (a field synthesizing philosophy of science, history of science, sociology of science, and philosohpy of technology)}}
	\acro{TAXII}{Trusted Automated eXchange of Indicator Information\acroextra{ (by \acs{MITRE})}}
    	\acro{TCP/IP}{Transmission Control Protocol / Internet Protocol}
    	\acro{TLA}{Temporal Logic of Actions}
    	\acro{TLP}{Traffic Light Protocol}
    	\acro{TLD}{Top-Level Domain\acroextra{ (in \acs{DNS})}}
	\acro{TSA}{Transport Security Administration\acroextra{ (\ac{US})}}
    	\acro{TTPs}{Tools, tactics, and procedures}
    	\acro{UN}{United Nations}
	\acro{UML}{Unified Modeling Language\acroextra{, see \citet{larman2001patterns}}}
	\acro{US}{United States of America}
    	\acro{USCERT}[US-CERT]{\ac{US} Computer Emergency Readiness Team\acroextra{, a branch of \acs{NCCIC} within \acs{DHS}}}
	\acro{URL}{Uniform Resource Locator}
	\acro{VERIS}{Vocabulary for Event Recording and Incident Sharing}
    	\acro{W3C}{World Wide Web Consortium}
    	\acro{XCCDF}{Extensible Configuration Checklist Description Format\acroextra{ (by \acs{NIST})}}
    	\acro{XML}{Extensible Markup Language\acroextra{, a standard by \acs{W3C}}}

\end{acronym}

\end{document}